\documentclass[sigconf]{acmart}

\usepackage{algorithmic,algorithm}
\usepackage{graphicx}
\usepackage{url}
\usepackage{multirow}
\usepackage{array}
\usepackage{booktabs}
\usepackage{multicol}
\usepackage{makecell}
\usepackage{subfig}
\usepackage{threeparttable}

\usepackage{amsmath,amssymb,amsthm}  
  
\DeclareMathOperator*{\argmax}{arg\,max}

\AtBeginDocument{%
  \providecommand\BibTeX{{%
    \normalfont B\kern-0.5em{\scshape i\kern-0.25em b}\kern-0.8em\TeX}}}

\copyrightyear{2020} 
\acmYear{2020} 
\setcopyright{acmlicensed}
\acmConference[MSR '20]{17th International Conference on Mining Software Repositories}{October 5--6, 2020}{Seoul, Republic of Korea}
\acmBooktitle{17th International Conference on Mining Software Repositories (MSR '20), October 5--6, 2020, Seoul, Republic of Korea}
\acmPrice{15.00}
\acmDOI{10.1145/3379597.3387439}
\acmISBN{978-1-4503-7517-7/20/05}

\begin{document}

\title{From Innovations to Prospects: What Is Hidden Behind Cryptocurrencies?}

\author{Ang Jia}
\email{jiaang@stu.xjtu.edu.cn}
\affiliation{
	\institution{Xi’an Jiaotong University}
	\department{School of Cyber Science and Engineering}
}

\author{Ming Fan}
\email{mingfan@mail.xjtu.edu.cn}
\affiliation{
	\institution{Xi’an Jiaotong University}
	\department{School of Cyber Science and Engineering}
}
\author{Xi Xu}
\email{xx19960325@stu.xjtu.edu.cn}
\affiliation{
	\institution{Xi’an Jiaotong University}
	\department{School of Cyber Science and Engineering}
}

\author{Di Cui}
\email{cuidi@stu.xjtu.edu.cn}
\affiliation{
	\institution{Xi’an Jiaotong University}
	\department{School of Cyber Science and Engineering}
}
\author{Wenying Wei}
\email{wwyhphrq@163.com}
\affiliation{
	\institution{Xi’an Jiaotong University}
	\department{School of Cyber Science and Engineering}
}

\author{Zijiang Yang}
\email{zijiang.yang@wmich.edu}
\affiliation{
	\institution{Western Michigan University}
	\department{Department of Computer Science}
}

\author{Kai Ye}
\email{kaiye@xjtu.edu.cn}
\affiliation{
	\institution{Xi'an Jiaotong University}
	\department{School of Electronics and Information Engineering}
}

\author{Ting liu}
\email{tingliu@mail.xjtu.edu.cn}
\affiliation{
	\institution{Xi’an Jiaotong University}
	\department{School of Cyber Science and Engineering}
}

\begin{abstract}


The great influence of Bitcoin has promoted the rapid development of blockchain-based digital currencies, especially the altcoins, since 2013. 
However, most altcoins share similar source codes, resulting in concerns about code innovations.
In this paper, an empirical study on existing altcoins is carried out to offer a thorough understanding of various aspects associated with altcoin innovations. 
Firstly, we construct the dataset of altcoins, including source code repository, GitHub fork relation, and market capitalization (cap). Then, we analyze the altcoin innovations from the perspective of source code similarities. The results demonstrate that more than 85\% of altcoin repositories present high code similarities. Next, a temporal clustering algorithm is proposed to mine the inheritance relationship among various altcoins. The family pedigrees of altcoin are constructed, in which the altcoin presents similar evolution features as biology, such as power-law in family size, variety in family evolution, etc. Finally, we investigate the correlation between code innovations and market capitalization. Although we fail to predict the price of altcoins based on their code similarities, the results show that altcoins with higher innovations reflect better market prospects.

\end{abstract}


\maketitle

\section{Introduction}

The great influence of Bitcoin\footnote{https://bitcoin.org/} has promoted the rapid development of blockchain-based digital currencies (cryptocurrency) since 2013. There are three critical features for existing cryptocurrencies: 1) Large variety, in August 2019, the number of available cryptocurrencies is over 2300 and it is still growing quickly\cite{coinmarketcap}. 2) Fast  delivery, the number of ICOs\footnote{\url{https://en.wikipedia.org/wiki/Initial_coin_offering}} (initial coin offering) newly increased per month is more than 60 since 2018. 3) Huge price change, a drastic fluctuation occurred in Bitcoin when it reached a peak of more than 20000\$ on Dec 17, 2017, and fell to 3830.04\$ on Dec 30 2018\cite{Bitcoin}. These three features have raised concerns about the quality of the cryptocurrencies and the real value of them.

The cryptocurrency world has experienced a complicated evolution since its birth. Bitcoin is the first cryptocurrency and now accounts for nearly 70\% of the market\cite{percentage}. The later cryptocurrencies can be divided into two categories: Alternative Cryptocurrency Coins (Altcoins) and Tokens\cite{differentaltcoins}. Altcoins refer to the coins that are alternatives to Bitcoin. Usually, they have their own chains. However, tokens serve the blockchain and they are often built on top of another blockchain such as the Ethereum blockchain. In this study, we only focus on altcoins. According to the source of altcoins, they can be divided into two groups: the variant of Bitcoin and the cryptocurrency with self-developed blockchain\cite{differentaltcoins}. The majority of altcoins are variants (forks) of Bitcoin, built using Bitcoin's open-sourced, original protocol with changes to its underlying codes, therefore conceiving an entirely new coin with a different set of features. Other altcoins are not derived from Bitcoin's open-source protocol. Rather, they have created their own blockchain and protocol that supports their native currency.

To determine whether a cryptocurrency should be invested, one key point is its code innovation. According to a report\cite{ChineseReport} released in September 2018, there exist great similarities between altcoins, i.e., \emph{``the code underpinning more than 90\% of the projects shared a similarity score of at least 80\%''}. The severity of code clone raises concerns about the innovation of altcoins. However, investors only focus on market trends and ignore the implementation details of the codes. Moreover, existing studies that analyze the cryptocurrency only focus on the technical issues, and they would ignore the code innovation. In addition, what is amazing for technologists is that the value of a technology product seems to have nothing to do with the technology's most direct vehicle, i.e., the code.  Therefore, it is non-trivial to get an understanding of the altcoin innovations, so well as their relation with the market prospects.

In this work, we aim to offer a thorough understanding of the various aspects associated with innovations of altcoins. 
Specifically, we first construct our dataset by crawling three different types of data from GitHub and CoinMarketCap\cite{coinmarketcap}, including source code repositories, GitHub fork relations of repositories and market capitalization (market cap) of altcoins. To conduct a thorough analysis, we empirically study the altcoin codes by answering three research questions listed below.

\emph{\textbf{RQ1}: To what extent the codes are shared by altcoins?}

\textbf{Innovations.} Many altcoins describe themselves with identical and innovative features and the ability to serve as better alternatives to Bitcoin. This ability actually lies in innovations of their implementation, i.e., source codes. However, it remains unknown what extent they share codes with other existing cryptocurrencies. Therefore, at the first step, we would like to offer an overview of the duplication in altcoin codes implemented in C++ (the largest part) to answer RQ1.


\emph{\textbf{RQ2:} What is the relation between altcoins?}

\textbf{Relations.} Software developments mostly do not start from scratch. Altcoins may also stem from one or many templates. Therefore, many altcoins may share similar source codes. However, the similarities between them cannot reveal the process of their inheritance. Although existing clustering algorithms offer clustering analysis of similar individuals, they ignore the sequence of input instances and cannot be applied in our work directly. In the second step, we aim to conduct the familial analysis of altcoins to further explore their inner relation by answering RQ2.

\emph{\textbf{RQ3:} What is the correlation between prospects and innovations?}

\textbf{Prospects.} Same with rules of nature that similar organisms will compete for the same resources, altcoins implemented with the same codes are more likely to suffer the same competitions. Therefore, innovations may play an important role in affecting the prospects of altcoins. However, the complicated environment in the cryptocurrency world makes it hard to understand it directly. At the last step, we are willing to offer a deep insight into the correlation between prospects and innovations by answering RQ3.


\textbf{Contributions:} Our main contributions are list below:

\renewcommand\theenumi{\roman{enumi}}
\renewcommand\labelenumi{(\theenumi)}
\begin{enumerate}
    
    
    
    
\item To the best of our knowledge, our empirical study for the first time investigated the innovations of altcoins from the view of source code, and we proposed a novel temporal clustering algorithm to mine the inner relations between altcoins by constructing family pedigrees.

\item We investigate the correlations between the altcoin innovations and their market capitalization from the view of code similarity. The results show that the altcoins with high innovations reflect better market prospects and, the cryptocurrency world is moving toward diversity.
 
\item We release the dataset of altcoin including the source code repositories, GitHub fork relations, and market caps. The dataset can be accessed from GitHub\footnote{https://github.com/island255/Cryptocurrencies-Coins-Sourcecode-download \\ https://github.com/island255/GitHub-Fork-relation \\https://github.com/island255/CrawlPrice}.

\end{enumerate}

\textbf{Overview of the paper:} Section \ref{sec:motivating} presents an motivating example. Section \ref{sec:method} presents the construction of our dataset and its statistics. Section \ref{sec:results} introduces the study design and results by answering the three research questions. After the discussing of the related work and threats to validity in Section \ref{sec:related work} and Section \ref{sec:threats}, we conclude our paper in Section \ref{sec:conclusion}.

\section{Motivating Example}
\label{sec:motivating}

This section motivates our research using a real-world example.  StrikeBitClub\footnote{\url{https://coinmarketcap.com/currencies/strikebitclub/}} and  HollyWoodCoin\footnote{\url{https://coinmarketcap.com/currencies/hollywoodcoin/}} are two altcoins, where Strike-BitClub is a hybrid scrypt PoW + PoS based one\footnote{\url{https://github.com/sbccoin/sbccoin-source}} and HollyWoodCoin is a PoS-based one\footnote{\url{https://github.com/hollywoodcoin-project/HollyWoodCoin}}. Table \ref{tab2} shows the collected information of this two cryptocurrencies. They are both released in December 2017, they both suffer a huge price loss from December 2017 to September 2018.

\begin{table}[htbp]
    \vspace{-10pt}
    \caption{StrikeBitClub and HollyWoodCoin}
    \begin{center}
        \begin{tabular}{|c|c|c|c|}
            \hline
            currency &  StrikeBitClub & HollyWoodCoin\\
            \hline
            release time &  2017/12/9 & 2017/12/14 \\
            \hline
            price(release) &  0.570882\$ & 17.62\$ \\
            \hline
            price(2018/4) & 0.008867\$ & 2.52\$ \\
            \hline
            status(2019/8) &  dead & dead \\
            \hline
        \end{tabular}
        \label{tab2}
        \vspace{-10pt}
    \end{center}
\end{table}

The source code repositories of these two altcoins and their GitHub fork information are crawled on 28 March 2018. They do not indicate GitHub fork relation between their repositories or to other repositories. However, using textual code clone detection, the two repositories share about 97\% of codes.

The `coincidences' in their codes, their release times, and their price loss inspire us to carry out studies to further mine the hidden phenomenons. Altcoins are running based on similar techniques, and these two altcoins share highly similar source codes, which inspire us to explore the innovations of other altcoins and the relation between these altcoins. Companies with the same technology in the same industry will compete for the same resources and these two similar altcoins suffer a decline in the price, which raises our exploration about the relation between code innovations and market prospects in the cryptocurrency world. To illustrate our approaches to solve these questions, we introduce our dataset in section \ref{sec:method}, and our study in section \ref{sec:results}.

\section{Dataset}
\label{sec:method}

This section presents details of the dataset. We crawl source code repositories, GitHub fork relations, and market caps of all altcoins from GitHub and CoinMarketCap.com to construct our dataset. The process of dataset construction and statistics of the dataset will be illustrated in the following subsections.

\subsection{Dataset Construction}
\textbf{Source Code Repositories:} By 4 April 2019, more than 3000 cryptocurrencies have emerged. Bitcoin has about 20,000 commits and more than 200 release versions. We collect all versions of the cryptocurrencies (including Coin and Token) to construct our dataset. Source code repositories of these cryptocurrencies were accessed through the following steps. First, we access the website of each cryptocurrency on CoinMarketCap.com and get their GitHub links. Then we access the GitHub links and identify whether it is a repository or repositories. If it is a repository, we will download this repository; otherwise, we will download the pinned repositories or (if the pinned repositories do not exist) the first three repositories.

\textbf{Market Caps:} We obtain the historical closing prices and the market capitalization of all cryptocurrencies by crawling the website CoinMarketCap.com. We use Node.js to crawl the information and store them in JSON format.

\textbf{GitHub Fork Information:} The distinction between GitHub fork and blockchain fork needs to be explained. A fork in the blockchain is when a change is made to the software of a cryptocurrency to create another version of the blockchain, technically\cite{Bitcoin-hardforks}, while A fork in GitHub is a copy of a repository. Bitcoin hard forks (blockchain fork) leads to the creation of many altcoins, such as Bitcoin Cash, Bitcoin Gold, etc. When an altcoin is a product of Bitcoin hard forks, there may be GitHub fork relation between their repositories. We would like to collect their GitHub fork relation between all altcoins before analyzing their code similarity relation. We query the GitHub API to obtain information of their repositories. Results are structured in JSON format, and the fork relation can be obtained from its ``parent" node.

\subsection{Statistics of Dataset}

By 4 April 2019, there come up 3338 cryptocurrencies (including coin and token). Among them, 1253 have opened their code links and 1230 cryptocurrencies set GitHub repositories as their links. Using the method in Section 3.1, we successfully access 1698 repositories belonging to 1023 cryptocurrencies (some links are invalid or unable to open). Among them, more than 40\% (693 in 1698) have only one release and the repository \emph{bitcoin/ bitcoin} of Bitcoin has the largest number of releases as 216 times. 

\begin{figure}[t]
    \centering
    \includegraphics[width=0.45\textwidth]{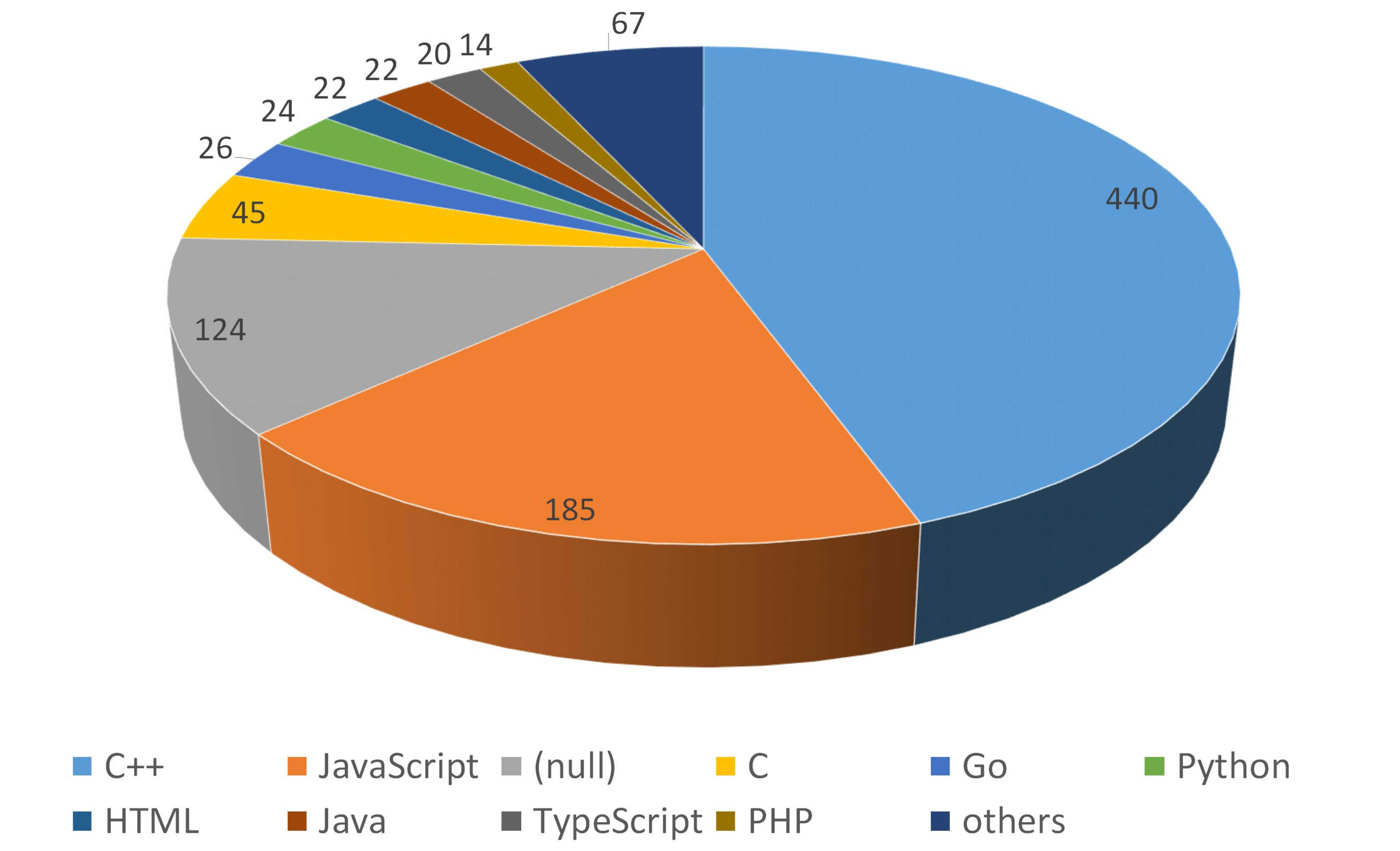}
    \caption{Top 9 language in cryptocurrencies' repositories}
    \label{fig:language}
    \vspace{-10pt}
\end{figure}

We try to crawl the information (created time, language, and fork) of these 1698 repositories using GitHub API, but only access the information of 989 repositories successfully. Among these 989 repositories, 865 repositories are labeled with a specific language. Figure \ref{fig:language} shows the proportion of languages used in these repositories.  Among the labeled 865 repositories, 440 repositories are labeled that its primary language is C++, and 185 repositories' main language is JavaScript. Other languages used in more than 20 repositories are C, Go, HTML, Java, Python, and TypeScript. 

\begin{figure*}
    \centering
    \includegraphics[width=0.7\textheight]{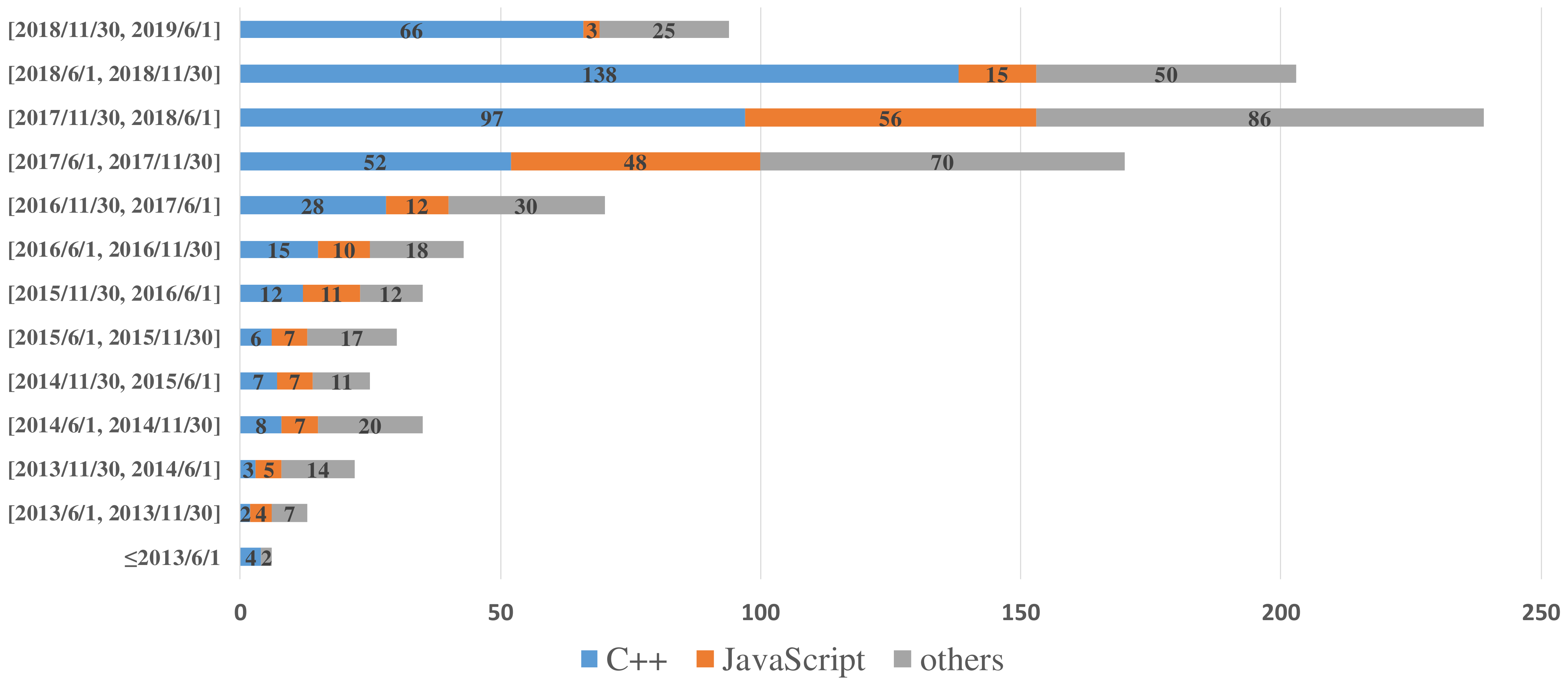}
    \caption{Number of repositories written in different languages during different periods}
    \label{fig:languagechange}
    \vspace{-10pt}
\end{figure*}

We also analyze their created times from our obtained statistics. Figure \ref{fig:languagechange} shows the number of repositories written in different languages during different periods. It indicates enthusiasm for creating repositories (including coin projects and token projects) since 2017. While the creating of all repositories arrives at its peak during 2017/11/30 to 2018/6/1, creating of C++ written repositories reaches its peak during 2019/6/1 to 2018/11/30. Moreover, from 2017/11/30 to 2018/6/1, there arise 239 repositories, which means a new repository comes out every 1.5 days.

\begin{figure}
    \centering
    \includegraphics[width=0.48\textwidth]{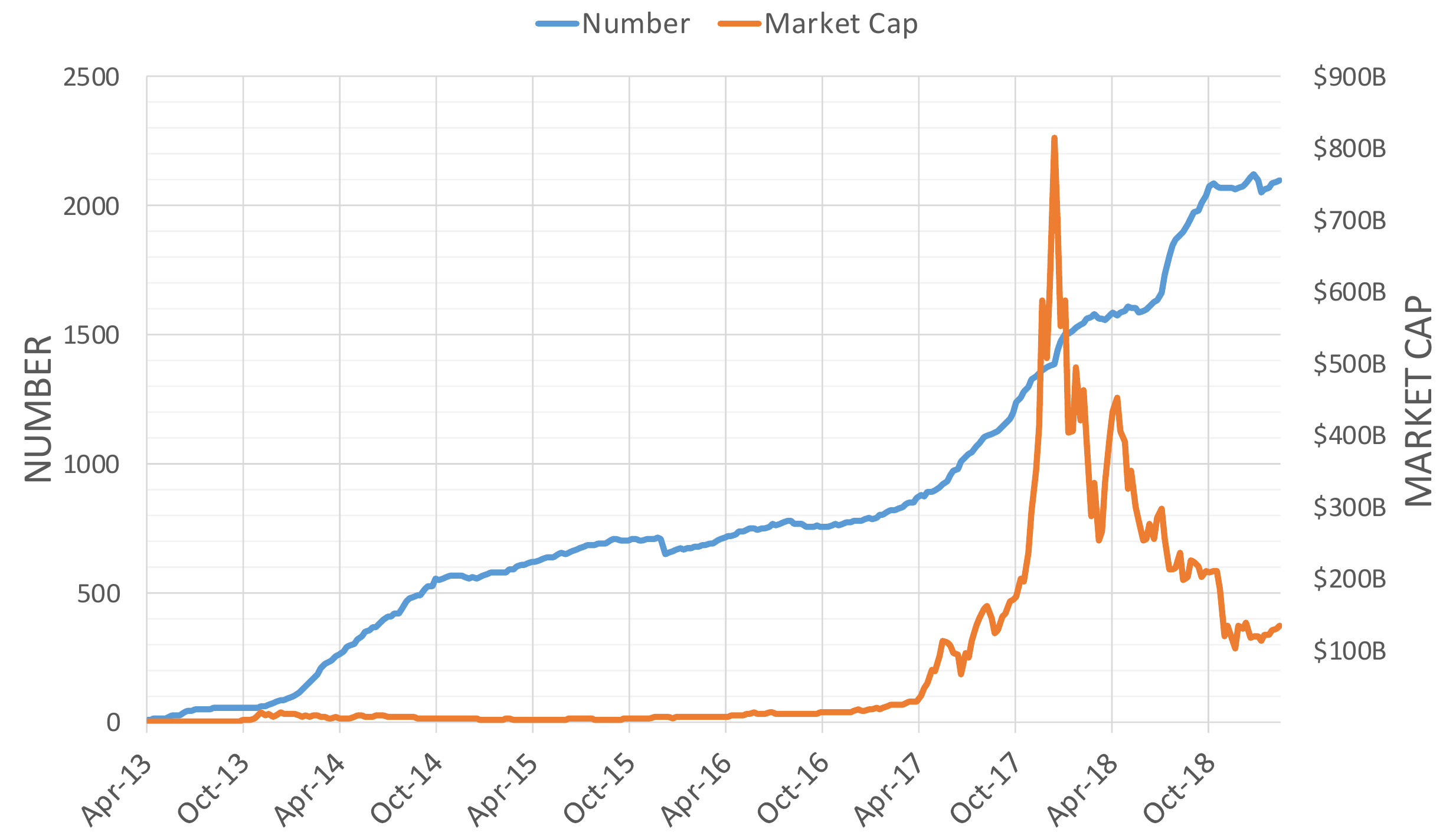}
    \caption{Total market capitalization and number of newly-released cryptocurrencies}
    \label{fig:Marketcap}
    \vspace{-10pt}
\end{figure}

Figure \ref{fig:Marketcap} shows changes in total market capitalization and the number of cryptocurrencies. The market capitalization of cryptocurrency experience a pull-up from 2017/1 to 2018/1 and a plunge after 2018/1. Besides, the number of newly-created repositories indicates a similar trend with a slight delay. Although cryptocurrencies spring up like mushrooms after the rain of market expansion, there may exist some hidden troubles in the course of this explosive development.

\section{Study Design and Results}
\label{sec:results}

\begin{figure*}[t]
    \centering
    \includegraphics[width=150mm]{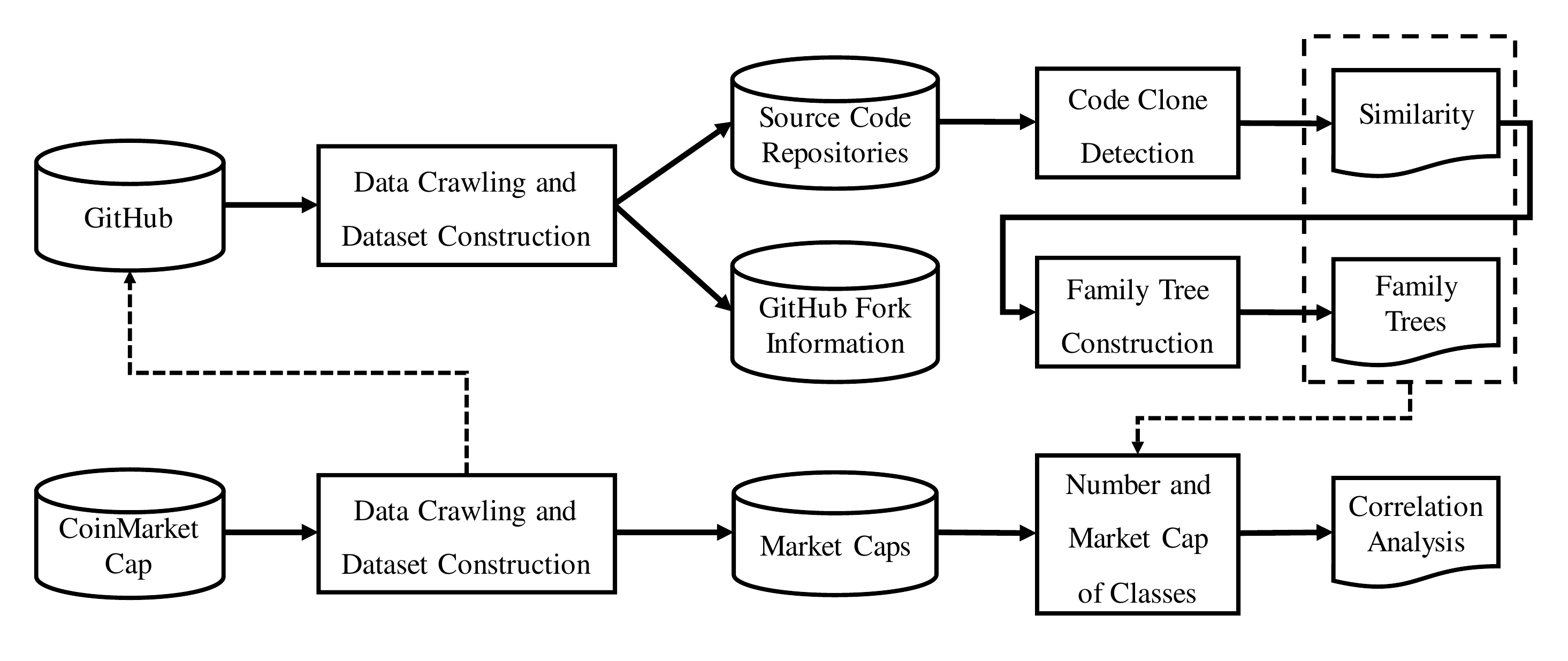}\\
    \caption{Overview of our approach}
    \label{fig:overvierw}
    \vspace{-5pt}
\end{figure*}

The overview of our study is presented in figure \ref{fig:overvierw}. First, we acquire market caps and source code repositories of different cryptocurrencies from CoinMarketCap.com and GitHub. Then, we select repositories written in C++ (the largest part) and then calculate the similarities between them. Next, we propose a novel clustering algorithm combined with temporality to construct families of cryptocurrencies based on code similarities. Finally, we compare the market prospects with their code similarities and relations. 

We present the study results by answering three research questions. For each question, we present the motivation behind the research question, the approach, and a discussion of our findings. 

\subsection{To what extent the codes are shared by altcoins? }
\label{q:1}

\textbf{Motivation:} Altcoins serve for a common purpose---exchange and demonstrate their features in offer, decentralization, immutability, anonymity, etc. However, even though an altcoin describes it has many innovative and meaningful creations in its new release, there may be little difference in its source code and others. To verify the true innovations of an altcoin, the analysis of their codes and the similarities with others are needed. 

\textbf{Approach:} Our study measures the global situation of cryptocurrencies' innovation from two aspects. First, we obtain their fork relation through the GitHub API. The GitHub API provides endpoints for users to consume GitHub data as well as make changes on a user's behalf\cite{Github_API}. We send requests through GitHub API and obtain a thorough description of this repository. With the query results structured in JSON, we are able to access the fork related information of a repository from the `parent' (directly fork) and `source' (original source) nodes.  We finally crawl the fork information of 1698 repositories belonging to 1023 cryptocurrencies deployed in GitHub. 


Second, we implement code clone detection to calculate the similarities between repositories of different cryptocurrencies. It is unnecessary and difficult to calculate the similarities between all repositories of every version belonging to all cryptocurrencies. We cut two snapshots from our dataset on 28 March 2018 (D1) and 28 September 2018 (D2). In each snapshot, we select repositories of altcoins of their latest version at that time. As comparing repositories written in different languages is out of our scope and the C/C++ written repositories account for the largest (see Figure \ref{fig:language}), the repositories whose main language is C/C++ are preserved. Table \ref{tab:two datasets} shows the statistics about these C/C++ written projects of these two snapshots. Time, number of altcoins, and minimum, maximum, and average of LOCs (lines of code) are presented in this table.

\begin{table}[htbp]
    \caption{Statistics for projects written in C++}
    \begin{center}
        \begin{tabular}{c|c|c|c|c|c}
            \hline
            Dataset & Time & Number & MinLOC & MaxLOC & AvgLOC \\
            \hline
            D1      &  Mar  & 485      & 15804   & 392312 & 60937  \\
            \hline
            D2      &  Sep  & 566      & 13495   & 2560877 & 60443  \\
            \hline 
        \end{tabular}
        \label{tab:two datasets}
    \end{center}
\end{table}

To calculate the similarities between these repositories, we first remove configuration files, readme documents, and files implemented in other languages. Then we treat remaining source code files as text and use RKR-GST (Running-Karp-Rabin Greedy-String-Tilling \cite{karp1987efficient}) algorithm to find the same code snippets in two repositories. This algorithm, first carried out by Wise, uses the GST (greedy string tilling) algorithm, which aims to find similar fragments following the order from long to short and reduce the complexity of string extension matching by KR (Karp Rabin) algorithm. In this process, these similar code snippets are one-to-one mappings in two repositories. Finally, considering that even though a large repository contains codes of a small one, the similarity between them should not be high due to the difference in their size, we use the following formula to calculate the similarity between repository A and B:

\begin{equation}\label{}
similarity= \frac{2 \times \left |A \bigcap B \right |}{\left | A \right |+\left | B \right |}
\end{equation}

We regard the codes of A and B as two sets. $\left |A \bigcap B \right |$ represents the size of matched code snippets between A and B, $\left | A \right |$ and $\left | B \right |$ respectively represent the size of A and B. The size can be measured by the length of character. 

For cryptocurrencies with one repository, we set similarities of this one repository with others as similarities of this altcoin. For cryptocurrencies with more than one repository, we merge the similarities of repositories by selecting the highest similarities. 


\textbf{Findings:} \textbf{About 15\% of repositories shows fork relation with other repositories.} We crawl the fork information of 1698 repositories belonging to 1023 cryptocurrencies deployed in GitHub and successfully get the information of 989 repositories belonging to 983 cryptocurrencies. In this 989 repositories, 146 repositories have a fork relation with other repositories. Among them, \emph{bitcoin/bitcoin}\footnote{\url{https://github.com/bitcoin/bitcoin}} is the most forked repository forked by 12 other cryptocurrencies. The second is \emph{PIVX-Project/PIVX}\footnote{\url{https://github.com/PIVX-Project/PIVX/}} which is forked by 7 cryptocurrencies. Besides, \emph{bitcoin/bitcoin} is also the most sourced repository. It is the source of 34 repositories, indicating that apart from 12 repositories which directly fork this repository, the other 22 repositories forking other repositories also come from \emph{bitcoin/bitcoin}. From this point, the implementation of Bitcoin does have a great influence on other cryptocurrencies.

\begin{figure}
    \centering
    \includegraphics[width=0.48\textwidth]{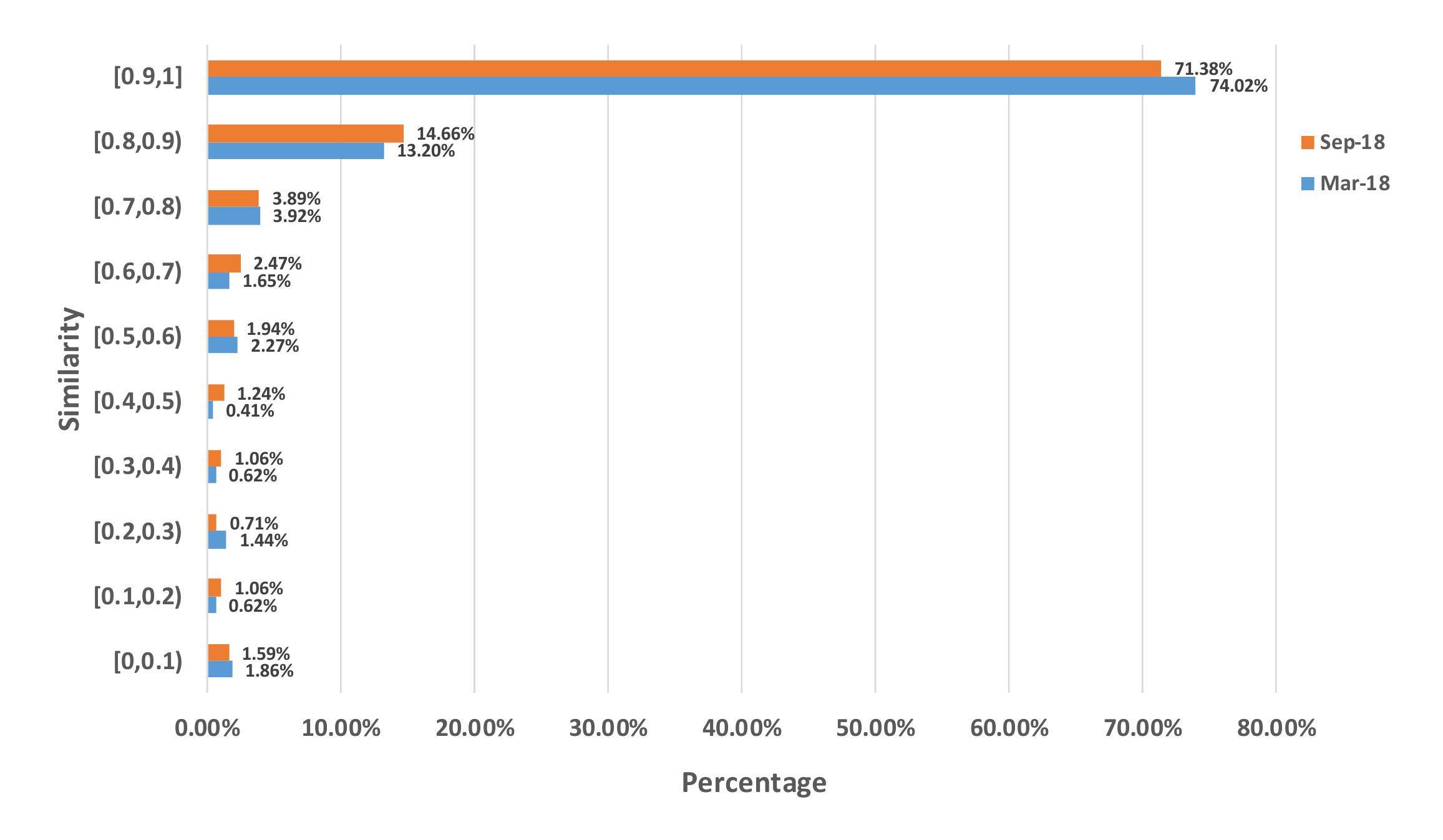}
    \caption{Distribution of maximum similarities with existed repositories}
    \label{fig:SimDistribution}
    \vspace{-10pt}
\end{figure}

\textbf{More than 85\% of the repositories share 80\% of codes with the other existing repositories.} Figure \ref{fig:SimDistribution} shows the distribution of maximum similarity of a repository with other existing repositories appearing before it. The similarities of these two experiments have a similar distribution that similarities between 90\% and 100\% account for the largest part, and the second are similarities between 80\% and 90\%. 

\textbf{33.92\% cryptocurrencies have a similarity of 80\% with more than 50 cryptocurrencies.} According to the results of D2, five cryptocurrencies have a similarity of more than 90\% with Bitcoin and 12 of more than 80\%. However, the number of cryptocurrencies whose similarities with PIVX more than 90\% and 80\% is 11 and 48. Besides, there are 55 cryptocurrencies of which each has a similarity of more than 90\% with more than 50 cryptocurrencies. Moreover, there are 192 cryptocurrencies (33.92\%) of which each also has a similarity of more than 80\% with more than 50 cryptocurrencies. 

\begin{center}
\fbox{\shortstack[l]{While few repositories show fork relation with others,\\ more than 85\% of the repositories share 80\% of codes \\ with the other earlier-created repositories. }}
\end{center}

\subsection{What is the relation between altcoins?}
\label{q:2}

\textbf{Motivation:} Conducting a thorough analysis of the relationship between different cryptocurrencies is critical. However, while the statistics show a high degree of duplication in codes of cryptocurrencies, it's hard to get a deeper understanding about their extended connections such as the vulnerability caused by their shared code snippets, the technology trends distributed in different groups and the inheritance of their codes during software development. In our study, we propose a temporal clustering algorithm to construct family pedigrees of cryptocurrencies.

\textbf{Approach:} To gain a deep insight into the relationship between cryptocurrencies, clustering them by similarities is a shortcut to acquiring an analyzable representation. However, existing clustering methods are insufficient to reflect the development of cryptocurrencies as they do not take temporality into account. We propose a novel clustering algorithm to present their similarities in chronological order.

Before the description of the algorithm, we introduce two rules to simplify the relationship between cryptocurrencies:

\emph{Rule 1. When the similarity between two cryptocurrencies is high, it is more likely to find that later-released cryptocurrency follows early-released cryptocurrency, which means the similarity is more like a directed relation.} 

\emph{Rule 2. When a cryptocurrency has high similarities with more than one cryptocurrency released early than it, it is more likely to follow the cryptocurrency with maximum similarity.} 

Based on the aforementioned rules, we only pay attention to pairs of altcoins which is most similar. Besides, we distinguish the different relation of two cryptocurrencies with different occurrence time intervals. With long-time intervals, we consider one cryptocurrency is a follower of the other one. With short time intervals, we consider they are in the same group. Finally, we propose our algorithm, as shown in algorithm \ref{alg:1}.

\begin{algorithm}
    \renewcommand{\algorithmicrequire}{\textbf{Input:}}
    \renewcommand{\algorithmicensure}{\textbf{Output:}}
    \caption{The Temporal Clustering Algorithm}
    \label{alg:1}
    \begin{algorithmic}[1]
        \REQUIRE altcoin[n], time[n], similarity[n][n] \\
        \COMMENT{n is the number of altcoins and altcoin[n],time[n] are sorted in chronological order}
        
        \ENSURE Family-Pedigrees
                
        \STATE $i=1$;
            
        \WHILE {$i<=n$ }
    
        \STATE put altcoin[i] in Family-Pedigrees;
        \STATE $ MaxSim[i] = \max \limits_{1 \leq x \leq n}similarity[x][i] $;
        \IF { i==1 or $ MaxSim[i] < \Theta_s $}
        \STATE set altcoin[i] as a initial node of a new tree;
        \STATE continue;
        \ENDIF

        \STATE $j =  \argmax \limits_{1 \leq j \leq n}similarity[j][i] $
        
        \IF {node[j] exists}
        \IF { $ \left | t(node[i])-t([node[x]]) \right |  > \Theta_t $}
        \STATE altcoin[i].fathercoin = altcoin[j];
        \ELSE
        \STATE altcoin[i].brothercoin = altcoin[j];
        \ENDIF
        \ENDIF
        \ENDWHILE

    \end{algorithmic}
\end{algorithm}

Algorithm \ref{alg:1} illustrates a way to construct family pedigrees for cryptocurrencies. At first, we sort cryptocurrencies by the chronological order of their occurrence time for convenience. Then, in line 2-8 in Algorithm \ref{alg:1}, we define that a new tree is created as a coin is selected without high similarities with early-released cryptocurrencies. Next, in line 9-16, we classify the relation between cryptocurrencies with maximum similarity based on the interval of their occurrence time. Finally, a forest composed of many family pedigrees will form. 

\begin{figure}
    \centering
    \includegraphics[width=0.39\textheight]{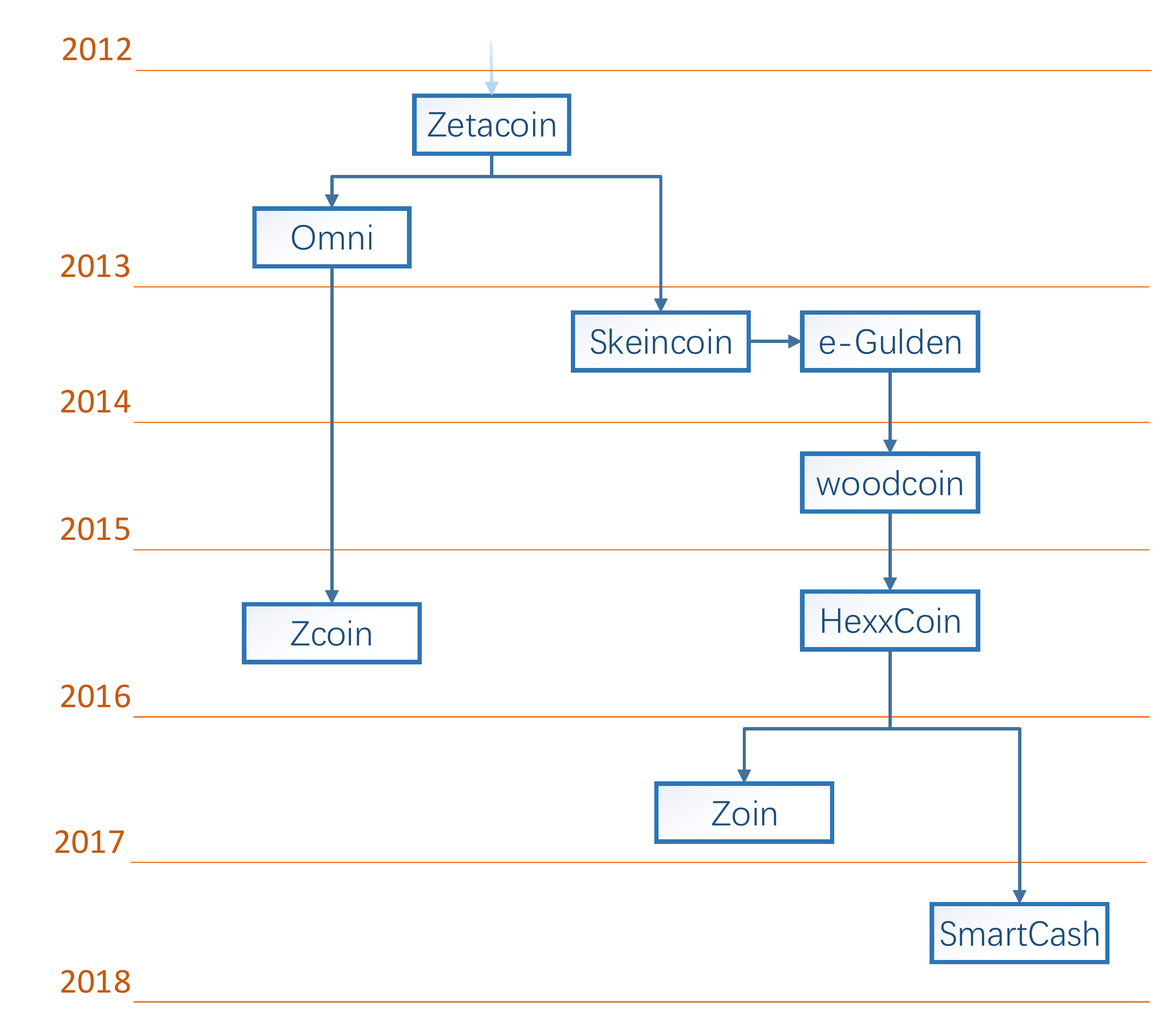}
    \caption{Part of a family pedigree in altcoin pedigrees}
    \label{fig:familytree}
\end{figure}

To have a graphic impression about the relations between hundreds of cryptocurrencies, we propose a way to present them in a tree diagram. The father coin will be the father node of this coin, while its brother coin will be in the same horizontal position. Figure \ref{fig:familytree} shows part of one family pedigree as our graphic representation of D2 (setting $\Theta_s=70\%$ and $\Theta_t =$ 3 months).  A horizontal arrow represents brotherhood, and an upright arrow indicates a father-son relationship. For example, \emph{Zetacoin} is the father node of \emph{Skeincoin} and \emph{e-Gulden} is the brother node of \emph{Skeincoin}.

\textbf{Findings:} \textbf{Most altcoins congregate in few groups.} We use our proposed algorithm to construct family pedigrees in D2 by setting $\Theta_s=70\%$. Figure 7 shows the distribution of the numbers in different families. The distribution has a big head and a long tail, which indicates an unbalanced distribution of altcoins in families. The phenomenon of this power-law distribution in altcoins is actually similar to the species richness in clades of organisms\cite{historical}. To be accurate, 72.4\% (410 in 566) of altcoins congregate in 8.8\% (5 in 57) of family pedigrees. It obeys the 80/20 principle, which indicates a more tight aggregation in the more technically-close groups.

\begin{figure}
    \centering
    \includegraphics[width=0.5\textwidth]{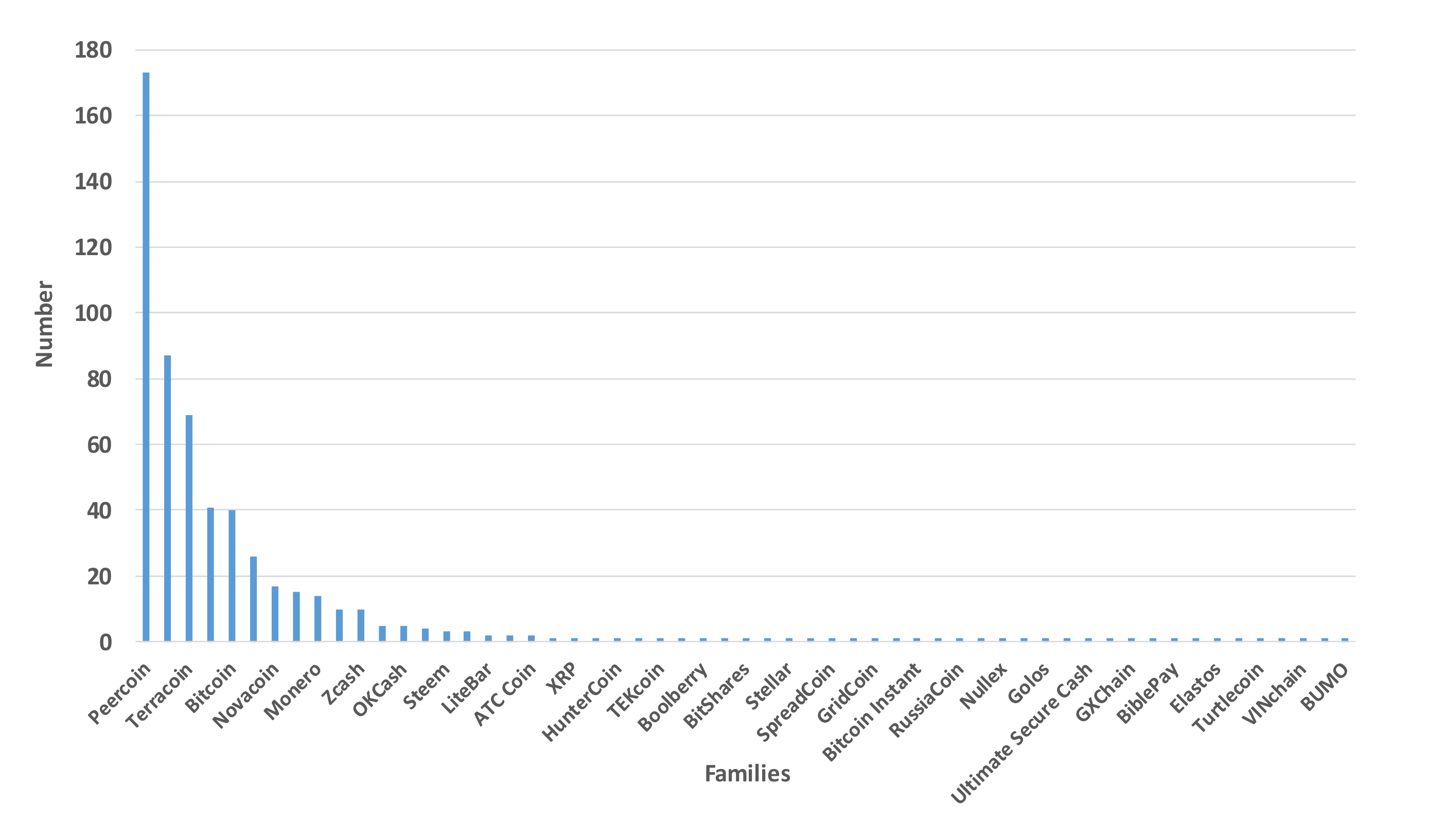}
    \caption{Number of altcoins in different families}
    \label{fig:familyDistribution}
    \vspace{-10pt}
\end{figure}

\begin{figure}
    \centering
    \includegraphics[width=0.47\textwidth]{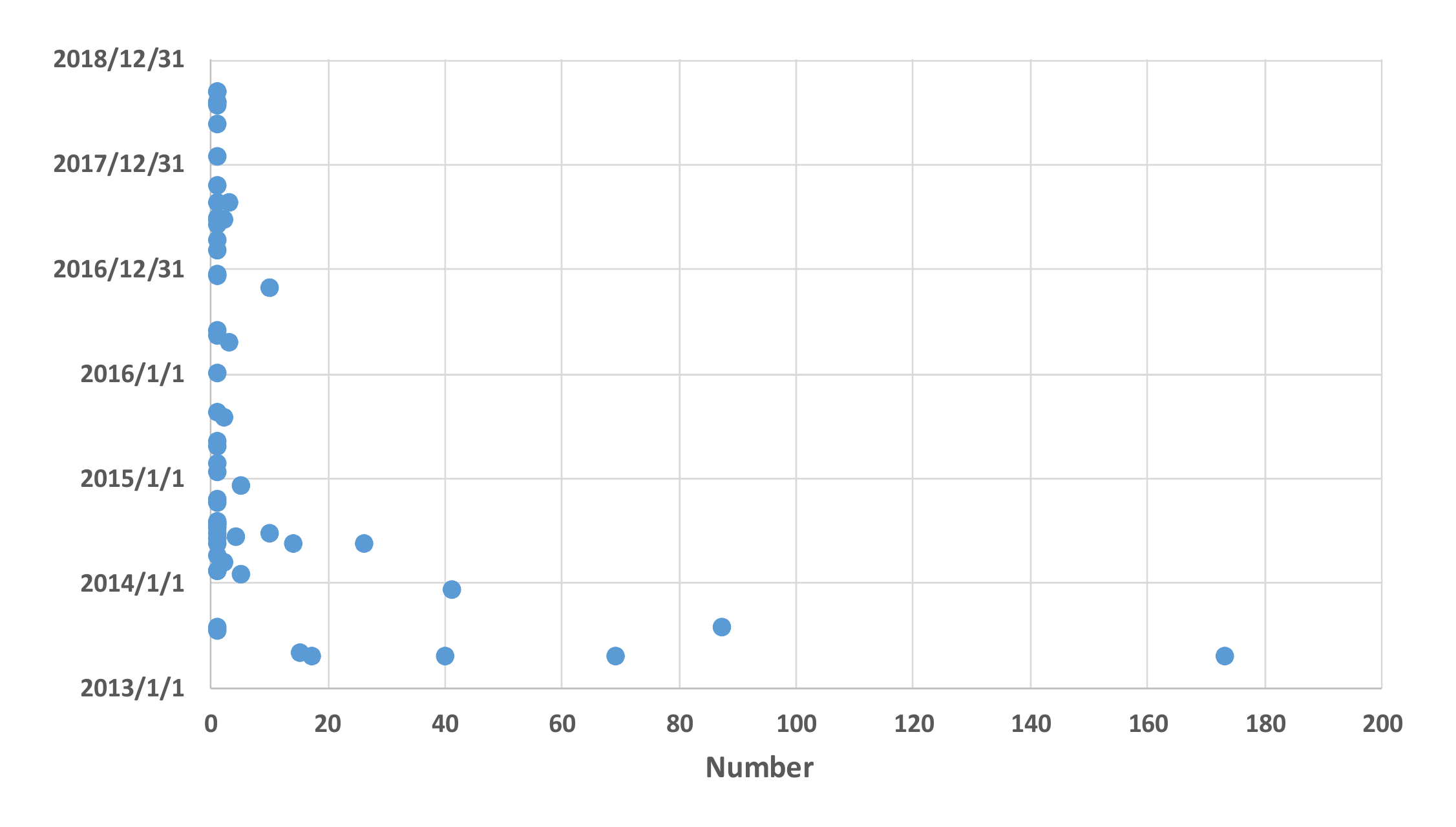}
    \caption{Time of the first altcoin in families trees}
    \label{fig:familiesandtime}
    \vspace{-10pt}
\end{figure}

\textbf{Early-released altcoins are more likely to form large families.} Figure \ref{fig:familiesandtime} shows the time of the first-released altcoin in families. Every node represents a family pedigree. The horizontal axis shows the number of altcoins in this family, while the vertical axis shows the release time of the first altcoin in this family. There are eight families with their first altcoins released before 2014, and 5 of them have more than 40 altcoins. There are 49 families of which altcoins are released after 2014, and 48 in them are families of less than 20 altcoins. There indicates a trend of families with early-released altcoins to own more members likely. We attribute this phenomenon to two reasons. One is that families with early-released altcoins will grow larger by our proposed algorithm. Algorithm \ref{alg:1} illustrates a way to construct family pedigrees from early-released altcoins. As a consequence, early-released altcoins will stay at a higher level and thus have more descendants. Another is that early-released altcoins have more reputations than later-released altcoins. The first several altcoins serve as a prototype for later ones, and Bitcoin, in particular, has a massive impact on the formation of other altcoins.

\begin{center}
    \fbox{\shortstack[l]{Most altcoins congregate in few families and early-\\  released altcoins are more likely to form large families.}}
\end{center}

\subsection{What is the correlation between prospects and innovations? }

\textbf{Motivation:} Charles Darwin has observed that due to the limited resources, there is a struggle for existence among individuals - often with only a fraction of offspring surviving through each generation to reproduce successfully. In the cryptocurrency world, there is also a rat race between different kinds of cryptocurrencies. Many factors may influence the prospect of a cryptocurrency, including but not limited to overall market trends, government policies, competitors with the same functionality, and its technology implementations. Especially, mining the relation between their technology implementations and their prospects are useful to see the importance of technology in the cryptocurrency market. 

In our previous study, we try to predict their price change from their codes (details in section 6.1). However, It is the same as common knowledge that the correlation between code similarity and market cap is weak. In this study, we carry out the following two questions to help further analyze their relation: 

\begin{itemize}
    \item[1.] What is the correlation between prospects of cryptocurrencies and code similarities?
    \item[2.] What is the correlation between prospects of cryptocurrencies and family pedigrees?
\end{itemize}

\textbf{Approach:} To solve question 1, we select D1 and D2 used in section \ref{q:1} as our research case. According to figure \ref{fig:Marketcap}, D1 and D2 are in a shrinking market with more competitive pressures which helps us to mine their correlation easier. To mine the correlation between prospects of cryptocurrencies and code similarities, we firstly define three classes of cryptocurrencies, which are cryptocurrencies without code link, cryptocurrencies with high similarities, and cryptocurrencies with low similarities. Then, to represents market prospects of these three classes, we define the rate of number change (NCR) and market cap change (MCCR) as follows:

\begin{equation}\label{}
    NCR(T)= \frac{Number(T_0 + T)}{Number(T_0)}
\end{equation}

\begin{equation}\label{}
    MCCR(T)= \frac{MarketCap(T_0 + T)}{MarketCap(T_0)}
\end{equation}

Prospects are calculated by analyzing their NCRs and MCCRs setting $T_0$ as half a year and a whole year. For D1, we calculate the change of their numbers and market caps from 28 March 2018 to 28 September 2018 and from 28 March 2018 to 28 March 2019. Similarly, we set the periods of D2 as from 28 September 2018 to 28 March 2019 and 28 September 2018 to 28 September 2019. 

From the prospects of different classes, we can discover some differences between cryptocurrencies with different similarities and cryptocurrencies without code links.

To solve question 2, we use family pedigrees (FP) of D1 and D2 generated by the algorithm \ref{alg:1} in section \ref{q:2}. First, we match the two sets of family pedigrees using algorithm \ref{alg:2}. As the maximum similarities (edges of family pedigrees) may change during this period, we only focus on the attributes of their nodes to establish their matches. In algorithm \ref{alg:2}, we match all pairs of FP1 and FP2 as long as there exist nodes in trees of FP1 appearing in trees of FP2. The consequence is that one tree in FP2 may match more than one tree in FP1. To avoid this embarrassment, we then select the match with most nodes in common. We search trees in FP1 (earlier) to match trees in FP2 (later) based on our assumption that one tree in the earlier family pedigree is likely to break up into two trees in the later family pedigree while two trees in the earlier one are not likely to unite into one tree in the later one. Then, same as the aforementioned approach, we calculate number change and prospects of family pedigrees from 28 March 2018 to 28 September 2018. From the statistics we processed, we present our research findings.


\begin{algorithm}
    \renewcommand{\algorithmicrequire}{\textbf{Input:}}
    \renewcommand{\algorithmicensure}{\textbf{Output:}}
    \caption{Family Pedigrees Matching Algorithm}
    \label{alg:2}
    \begin{algorithmic}[1]
        \REQUIRE Family-Padigrees1 (FP1), Family-Padigrees2 (FP2) \\
        \COMMENT{FP1 is earlier than FP2}
        
        \ENSURE Matches[ ]

        \STATE FP2 = Pruning(FP1, FP2); \\ 
        \COMMENT{Remove the nodes not in FP1 from FP2}

        \WHILE {trees in FP2 remain unvisited}
                
        \STATE Match = [ ];
        
        \STATE tree2 = GetTree(FP2); \\
        \COMMENT{take a family pedigree unviested from FP2}
        
        \STATE SetVisited(tree2);
        
        \WHILE {trees in FP1 remain unvisited}
        
        \STATE tree1 = GetTree(FP1);
        
        \STATE SetVisited(tree1);
        
        \IF {tree1 $\cap$ tree2 != $\emptyset$}
        \STATE Match $\leftarrow$ \{tree1,tree2\};
        \ENDIF
        
        \ENDWHILE
        \STATE Matches $\leftarrow \argmax \limits_{\{tree1,tree2\}\in match} \ \ \| tree1 \cap tree2\|$ ;
        
        \STATE SetUnvisited(FP2);
        
        \ENDWHILE

    \end{algorithmic}
\end{algorithm}

\begin{table*}[htbp]
    \caption{Number change and prospects of cryptocurrencies with different similarities}
    \begin{center}
        \renewcommand{\multirowsetup}{\centering}
        \begin{tabular}{c|c|c|c|c|c|c}
            \hline
            Snapshots  &  Classes & \makecell{Number/market cap \\at occurrence Time} & \makecell{Number/market cap\\ in half A Year(HA)} &  \makecell{Number/market cap\\ in one Year(WA)}  & \makecell{NCR/NCCR \\ rate (HA)} & \makecell{NCR/NCCR \\ rate (WA)} \\
            \hline
            \multirow{3}{*}{D1} & with code               & 644/2.26E+11\$
            & 514/1.99E+11\$               &453/1.20E+11\$      & 0.80/0.88  &0.70/0.53 \\
            \cline{2-7}
            & without code     & 223/5.02E+08\$
            & 143/4.63E+08\$                &115/2.66E+08\$    & 0.64/0.92  &0.52/0.53\\
            \cline{2-7}
            &  $ \ge 80\%$     & 408/1.04E+10\$
            & 313/6.34E+09\$               &268/5.24E+09\$      & 0.77/0.61  &0.66/0.50 \\
            \cline{2-7}
            &  $ \le 80\%$    & 60/1.80E+11\$
            & 51/1.67E+11\$               &46/1.03E+11\$      & 0.85/0.93  &0.77/0.57 \\
            \hline
            
            \multirow{3}{*}{D2} & with code               & 676/2.06E+11\$
            & 596/1.26E+11\$               &566/1.96E+11\$      & 0.88/0.61  &0.84/0.95 \\
            \cline{2-7}
            & without code     & 151/1.15E+09\$
            & 121/5.89E+08\$                &108/7.47E+08\$    & 0.80/0.51  &0.72/0.65\\
            \cline{2-7}
            &  $ \ge 80\%$    & 462/9.19E+09\$
            & 403/7.05E+09\$               &382/6.87E+09\$      & 0.87/0.77  &0.83/0.75 \\
            \cline{2-7}
            &  $ \le 80\%$    & 74/1.76E+11\$
            & 67/1.08E+11\$                       &64/1.80E+11\$      & 0.91/0.62  &0.86/1.02 \\
            \hline
        \end{tabular}
        \label{tab:differentsim}
    \end{center}
\end{table*}

\textbf{Findings:} \textbf{Altcoins with code links are more likely to have better prospects than altcoins without.} Table \ref{tab:differentsim} shows changes in the number and market cap of different classes of cryptocurrencies. We classify ``with code" and ``without code" through their websites on CoinMarketCap.com. When there is a "Source Code" link in its main page, we classify it as "with code". Otherwise, it will be classified as "without code". To be more accurate, we make our best effort to verify the altcoins without code link through searching on Google and GitHub. NCR/NCCR rate (HA/WA) in Table \ref{tab:differentsim} represents NCR and MCCR of this class in half a year or a whole year. From the statistics, we can see that the number change rates of altcoins with code links are all higher than those without, which indicates that altcoins with code implementation as their technical support are more likely to survive in a shrinking market. Besides, in most cases, the market cap rates of altcoins with code are higher than those without. All suggest that altcoins with code links are more likely to have better prospects.


\textbf{Altcoins with low similarities are more likely to have better prospects than altcoins with high similarities.} Table \ref{tab:differentsim} pre-sents an example by using 80\% as the threshold to distinguish altcoins with high similarities and altcoins with low similarities. The similarity of an altcoin is defined as its maximum similarity with earlier-released altcoins. The changes of number and market cap are also calculated in the same way with classes "code" and "without code". From the statistics, it may arise doubt that altcoins with low similarities have fewer members but own larger market caps. This is because Bitcoin dominates the market and is the first cryptocurrency. In all cases, altcoins with low similarities have higher NCRs than altcoins with high similarities. Also, in most cases, altcoins with low similarities have higher MCCRs than altcoins with high similarities. All suggest that altcoins with low similarities are more likely to have better prospects.

\begin{figure*}[t]
	\vspace{-10pt}
	\centering
	\subfloat[families in D1]
	{
		\begin{minipage}[t]{0.45\linewidth}
			\label{Fig-3}
			\centering
			\includegraphics[width=1\textwidth]{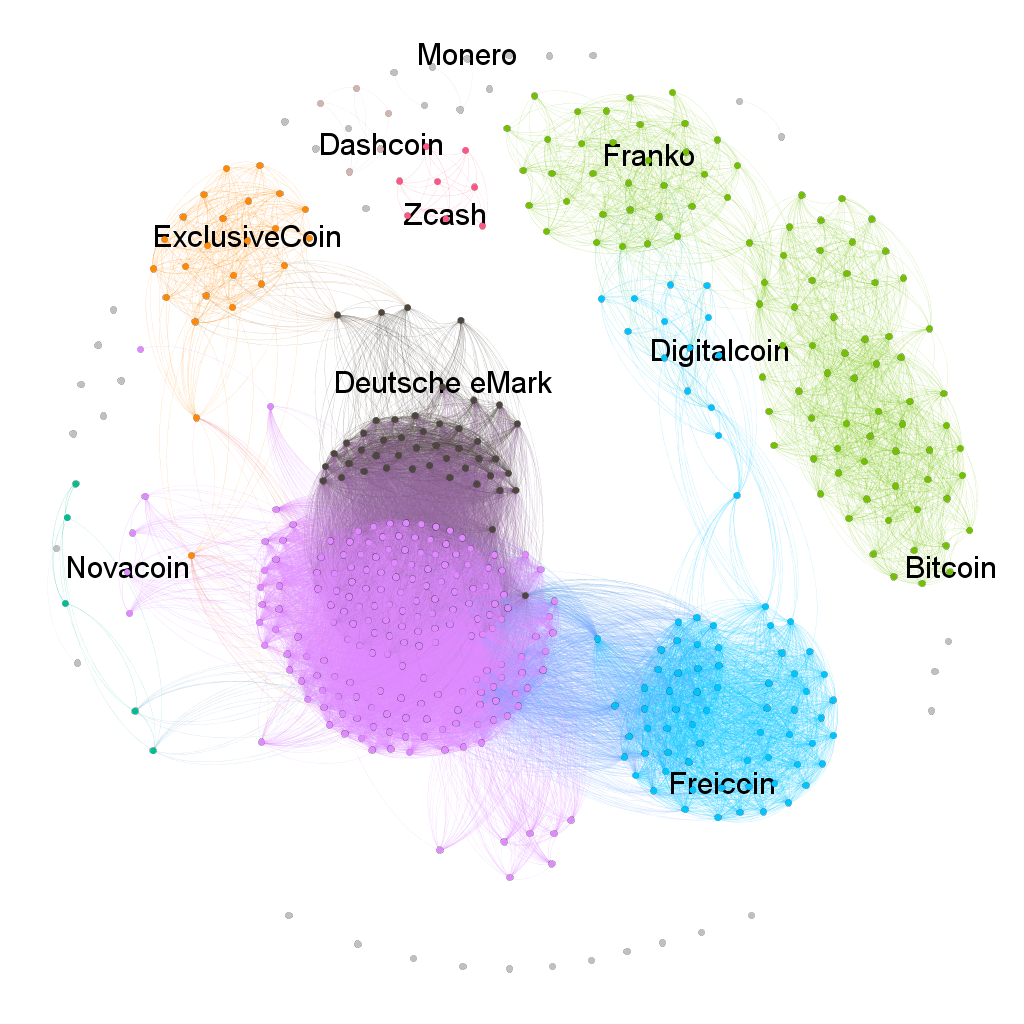}            
		\end{minipage}
	}
	\hspace{0mm}
	\subfloat[families in D2]{
		\begin{minipage}[t]{0.45\linewidth}
			\label{Fig-9}
			\centering
			\includegraphics[width=1\textwidth]{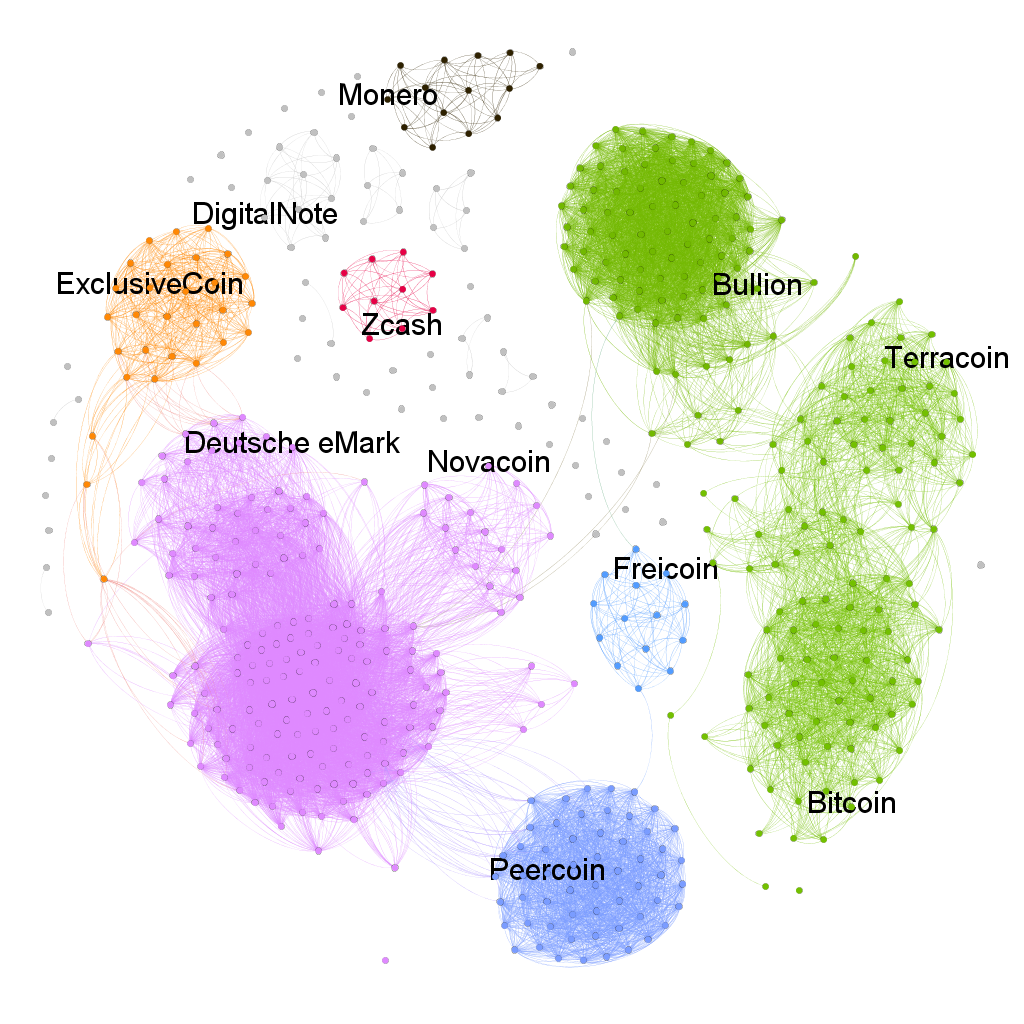}
		\end{minipage}
		
	}
	\caption{Families in D1 and D2}
	\label{fig:families}    
	\vspace{-10pt}
\end{figure*}

\textbf{Small families are likely to have better prospects than big families and the cryptocurrency world are moving toward diversity.} Figure \ref{fig:families} present graphical representation of families in D1 and D2. Table \ref{tab:differentfamily} shows details of the families comprised of more than ten altcoins. Families in the right (D2) indicate matches with families in the left (D1). Numbers and market caps are listed for each family, and the last column shows the number of altcoins shared by two families. Firstly, we found that the evolution of altcoins is similar to the evolution of species, such as one species may emerge, become extinct, or split into two species. For example, family \emph{Bitcoin} and \emph{Terracoin} in D2 all come from family \emph{Bitcoin} in D1, and they share 20 and 38 altcoins with family \emph{Bitcoin} in D1, separately. Besides, families of different size are experiencing different evolution. In the 11 families except for the second, families whose members are less than 40 experience an increase in their numbers, while families whose members are more than 40 experience a decrease both in their numbers and market caps. Although most families experience a loss in their market cap, families comprised of more altcoins seem to suffer more. In this way, small families are likely to have better prospects than big families. Due to the reduction of big families and the growth of small families, all families are moving to medium size. It is the same with biological evolution as this ecosystem is moving toward diversity.

\begin{table*}[htbp]
    \centering
    \caption{Number change and prospects of cryptocurrencies in different families}
    \begin{threeparttable}
        \begin{tabular}{c|c|c|c|c|c|c}
            \hline
            \multicolumn{3}{c|}{28 March 2018 (D1)} & \multicolumn{3}{c|}{28 September 2018 (D2)} & \multirow{2}{*}{$\left |D1 \bigcap D2 \right |$}  \\
            \cline{1-6}
            family* & number & market cap & family* & number & market cap &     \\
            \hline
            \multirow{2}{*}{Bitcoin**} & \multirow{2}{*}{63}    & \multirow{2}{*}{1.27E+11} & Bitcoin& 40     & 1.19E+11 & 20  \\
            \cline{4-7}
             &    &   &  Terracoin** & 69     & 2.85E+09 & 38  \\
            \hline    
            
            Dashcoi** & 5   &3.09E+06  & DigitalNote** & 10    & 2.85E+07 & 2   \\
            \hline
            Deutsche eMark  & 45  & 1.44E+08 & Deutsche eMark& 41     & 3.72E+07 & 30  \\
            \hline
            Digitalcoin** & 14  &1.28E+08 & Freicon** & 15     & 2.85E+09 & 10  \\
            \hline
            ExclusiveCoin   & 22  & 2.76E+08 & ExclusiveCoin & 26     & 1.21E+08 & 16 \\
            \hline
            Franko**  & 33  & 6.77E+08 & Bullion** & 87     & 2.54E+08 & 25  \\
            \hline

            Freicoin** & 238 &1.46E+09 & Peercoin**& 173     & 5.54E+08 & 147   \\
            \hline

            Monero & 3   &2.79E+09  & Monero & 14     & 2.11E+09 & 5   \\
            \hline
            Novacoin & 17   &7.74E+07  & Novacoin& 17     & 2.45E+07 & 14 \\
            \hline

            Zcash & 8   &7.42E+08 & Zcash & 10     & 8.06E+08  & 8   \\
            \hline
            single***  &29   &5.81E+10 & single*** & 38    & 3.43E+10  & 18   \\
            \hline
          
        \end{tabular}

        \begin{tablenotes}
            \footnotesize
            \item[*]We use the first coin of a family to represent that family. For example, \emph{Bitcoin} represents a family in which the first released coin is Bitcoin.
            
            \item[**]  As altcoins evolve from D1 to D2, the first altcoin of a certain family may change as some altcoin may die. Also, the code of a certain altcoin may change, too. So the presentation of a certain family and the member of a family will change according to their survival state and code similarity.
            
            \item[***] ``single'' represents families comprised of only one altcoin.
        \end{tablenotes}
         
    \end{threeparttable}
    \label{tab:differentfamily} 
\end{table*}

\begin{center}
    \fbox{\shortstack[l]{Altcoins with code links, low similarity, and in small \\ families are preferred, and the cryptocurrency world is \\ moving toward diversity. }}
\end{center}

\section{Related Work}
\label{sec:related work}

We discuss related works in two areas: analyses of cryptocurrencies and code clone detection.

\textbf{Analyses of Cryptocurrencies.} There are many researches that measure properties of cryptocurrencies from several aspects including but not limited to peer-to-peer networks \cite{biryukov2014deanonymisation, DBLP:journals/corr/MillerMLN17, DBLP:journals/cacm/MeiklejohnPJLMV16, DBLP:conf/fc/SpagnuoloMZ14}, data encryption \cite{DBLP:conf/fc/2014, DBLP:conf/socialcom/ReidH11}, attack \cite{apostolaki2017hijacking, DBLP:conf/uss/KapposYMM18,DBLP:conf/fc/VasekTM14} and scam identifications \cite{bartoletti2017dissecting, DBLP:conf/www/ChenZCNZZ18}. In this paper, we only focus on papers most related to our own. Reibel et al.\cite{DBLP:conf/fc/ReibelYM19} carried out research to identify the extent of innovation in the cryptocurrencies landscape using the open-source repositories. They focused on the code reuse of Bitcoin by using hash to calculate the similarities between repositories and analyze their relations through graphs. And they illustrated 8 clusters with directory structure similar relations and file similar relations. Azouvi et al. \cite{DBLP:conf/fc/AzouviMM18} conducted a study to analyze the centralization in the existing governance structures of Bitcoin and Ethereum. They collected discussions of the contributors and their contributions to their GitHub repositories and found that Ethereum appears more centralized than Bitcoin in terms of improvement proposals, but is more decentralized in terms of the discussion around its codebase. In our work, we further analyze the relationship between innovation and prospects and find many significant findings that can provide suggestions for investors and developers.

\textbf{Code Clone Detection.} Tools and techniques of code clone detection can be classified as six categories \cite{ain2019a}: textual approaches \cite{ragkhitwetsagul2017using, alhazmi2005quantitative, jadon2016code, yu2017detecting, kim2017vuddy, DBLP:conf/pldi/XueVL18, DBLP:conf/etfa/ThallerRPE17, DBLP:conf/icse/NewmanSCAM16}, lexical approaches \cite{nishi2018scalable, DBLP:journals/cee/TekchandaniBS17, DBLP:conf/icse/WangSWXR18, yuki2017a, sajnani2016sourcerercc, DBLP:conf/apsec/SemuraYCI17}, tree-based approaches \cite{yang2018structural, pati2017a, chodarev2015haskell}, metric-based approaches \cite{DBLP:conf/icse/SvajlenkoR17, DBLP:conf/icse/SvajlenkoR17a, DBLP:conf/iwsc/Ragkhitwetsagul18}, semantic approaches \cite{DBLP:conf/apsec/WangWX17, DBLP:conf/iwsc/SabiHK17, DBLP:journals/tmc/CrussellGC15, DBLP:conf/iwpc/HuZLG17, DBLP:conf/iwsc/KamalpriyaS17,fan2018android} and hybrid approaches \cite{roopam2017to, DBLP:conf/apsec/MisuS17, DBLP:conf/wcre/VislavskiRCB18, DBLP:conf/apsec/MisuSS17, DBLP:conf/compsac/AkramSML18, DBLP:journals/eswa/SheneamerRK18,fan2019ctdroid,fan2019graph}. In particular, we introduce several works closely related to our work. Nakamura et al. \cite{DBLP:conf/wcre/NakamuraCYHI16} introduce an approach to detect inter-language clones for web applications. As web applications contain source code written in co-dependent multiple programming languages, they merged co-dependent code clones detected from each programming language to further detect code clones. Yu et al. \cite{yu2017detecting} propose a multi granularity code clone detection method based on Java bytecode to detect code clones at both method level and block level. Apart from the similarity calculation between instruction sequences, they also calculate the similarity between the method call sequences to improve its effectiveness in detecting semantic code clones. Liu et al. \cite{liu2017vfdetect} propose VEDFECT, a system to detect vulnerable code clone based on fingerprint. The fingerprint is constructed by applying hash function to code blocks in the diff. Then the vulnerable code clone is detected by matching the preprocessed code blocks in the target project with the fingerprint. In our work, we further carry out analyses based on the code similarities acquired by code clone detection.

\section{Discussion and Threats}
\label{sec:threats}

\subsection{Discussion}
\textbf{Vulnerability inheritance.} Security is one of the most severe issues in cryptocurrencies. In 2018, Tim Ruffing \cite{ruffing2018burning} suggests that there was a cryptographic denial-of-spending attack on the original Zerocoin protocol, which is used in several cryptocurrencies including Zcoin, PIVX, SmartCash, Zoin, and HexxCoin. Although these cryptocurrencies handled this vulnerability through their ways, we believe that there may be some connections between their codes. As shown in Figure \ref{fig:familytree}, HexxCoin is the father node of Zoin and SmartCash, and Zcoin is also a member of this family. Although the similarity relation between cryptocurrencies is not equal to the inheritance of vulnerabilities, the family relation, which indicates a highly-similar relation, is useful for discovering hidden vulnerability associations to some extent. We did not carry out a strict process to examine the effectiveness of family pedigrees as it is not the focus of this research.

\textbf{Price prediction.} Interests are always what investors chase for in the cryptocurrency world. As we found high similar family pedigrees of altcoins, we wonder whether there is potential to predict their price change from their code similarities. Then we use the similarities of an altcoin with others as its feature, define the range of its relative price change against Bitcoin ($<0.1$, between 0.1 and 1, and $>1$) as its label, and use multiple machine learning models such as decision tree, Naive Bayes, and multi-layer neural network to predict the label of the newly-released altcoins by the model trained based on earlier-released altcoins. However, the accuracy shows a decline year by year, as the accuracy of predicting altcoins released in 2015 based on models trained using altcoins before 2015 is about 60\% while the accuracy of 2017 and 2018 is about 40\% and 33\%. Though it may be partly due to reasons caused by the change of dataset size, it reveals a hidden truth that the technical implementation of altcoins has less and less impact on price changes. 

\subsection{Threats}
\textbf{Dataset.} Our study focuses on altcoins written in C++. Altcoins implemented in other languages will be analyzed in our subsequent works. 
We select two snapshots on 28 March 2018 and 28 September 2018 to carry out our research. As the cryptocurrency world experienced a loss of market cap from 28 March 2018 to 28 September 2018, it would be more evident to see the difference of market prospects between different kinds of altcoins. But these two snapshots are limited, and we plan to give continuous analyses across several periods.

\textbf{Code clone detection.} We treat source code as text and use RKR-GST algorithm to calculate their similarities. It is simple, so it can't resist some obfuscations. But we hold the opinion that obfuscations are modifications to the source code, and we are supposed to take that difference into account. And we can get a thorough graphic representation of the same snippets to support the similarity results as the ground truth. When applying it to cryptocurrencies whose codes are written in other languages, it is also adequate for similarity calculation.

\textbf{Threshold.} We propose two thresholds as $\theta_s$ and $\theta_t$ to impose restrictions on family pedigree construction. In our results, we generate family pedigrees by setting $\Theta_s=70\%$ and $\Theta_t = 3 months$. And we use these results to analyze the relationship between prospects and families further. We also set 80\% as the threshold to distinguish cryptocurrencies with high similarities and low similarities. As we set the threshold by a certain value, the results may be a little different from the results obtained by other thresholds. However, the relation between innovations and prospects remains unchanged.

\textbf{Correlation between prospects and innovations.} We carry out a correlation analysis between codes and market caps in Section 4.3 and price prediction in Section 6.1, to investigate whether the code innovation of an altcoin is an essential factor of its market caps. It is the same as common knowledge that the correlation between code similarity and market cap is weak. However, our works show altcoins with high innovation tend to have better prospects, emphasizing the importance of innovation.

\section{Conclusion}
\label{sec:conclusion}

In this paper, an empirical study on existing altcoins is carried out to offer a thorough understanding of various aspects associated with altcoin innovations. 
Firstly, we construct the dataset of altcoins, including source code repositories, GitHub fork relations, and market capitalization (cap). Then, we analyze the altcoin innovations from the perspective of source code similarities. The results demonstrate that more than 85\% of altcoin repositories present high code similarities. Next, a temporal clustering algorithm is proposed to mine the inheritance relationship among various altcoins. The family pedigrees of altcoin are constructed, in which the altcoin presents similar evolution features as biology, such as power-law in family size, variety in family evolution, etc. Finally, we investigate the correlation between code innovations and market capitalization. Although we fail to predict the price of altcoins based on their code similarities, the results show that altcoins with higher innovations reflect better market prospects.

Our results emphasize the importance of code innovation and we suggest that newly-released cryptocurrencies pay more attention to technological innovations. Our work can be expended in several directions, such as the inheritance of code libraries, innovation analysis of tokens, relation between releases of highly-similar altcoins, and evolution of codes in the cryptocurrency world.

\newpage


\bibliographystyle{ACM-Reference-Format}
\bibliography{sample-base}


\begin{thebibliography}{60}


\ifx \showCODEN    \undefined \def \showCODEN     #1{\unskip}     \fi
\ifx \showDOI      \undefined \def \showDOI       #1{#1}\fi
\ifx \showISBNx    \undefined \def \showISBNx     #1{\unskip}     \fi
\ifx \showISBNxiii \undefined \def \showISBNxiii  #1{\unskip}     \fi
\ifx \showISSN     \undefined \def \showISSN      #1{\unskip}     \fi
\ifx \showLCCN     \undefined \def \showLCCN      #1{\unskip}     \fi
\ifx \shownote     \undefined \def \shownote      #1{#1}          \fi
\ifx \showarticletitle \undefined \def \showarticletitle #1{#1}   \fi
\ifx \showURL      \undefined \def \showURL       {\relax}        \fi
\providecommand\bibfield[2]{#2}
\providecommand\bibinfo[2]{#2}
\providecommand\natexlab[1]{#1}
\providecommand\showeprint[2][]{arXiv:#2}

\bibitem[\protect\citeauthoryear{??}{Chi}{2018}]%
        {ChineseReport}
 \bibinfo{year}{2018}\natexlab{}.
\newblock \bibinfo{title}{{Chinese report finds 9 in 10 altcoins to have stolen
  80\% of the code}}.
\newblock
  \bibinfo{howpublished}{\url{https://news.bitcoin.com/chinese-report-finds-9-in-10-altcoins-stolen-80-code}}.
\newblock
\newblock
\shownote{[Online; accessed 2-December-2019].}


\bibitem[\protect\citeauthoryear{??}{Bit}{2019a}]%
        {Bitcoin-hardforks}
 \bibinfo{year}{2019}\natexlab{a}.
\newblock \bibinfo{title}{{A Full List Of Bitcoin Hard Forks (UPDATED 2019)}}.
\newblock
  \bibinfo{howpublished}{\url{https://www.investinblockchain.com/a-full-list-of-bitcoin-hard-forks/}}.
\newblock
\newblock
\shownote{[Online; accessed 3-September-2019].}


\bibitem[\protect\citeauthoryear{??}{coi}{2019}]%
        {coinmarketcap}
 \bibinfo{year}{2019}\natexlab{}.
\newblock \bibinfo{title}{{All Cryptocurrencies | CoinMarketCap}}.
\newblock
  \bibinfo{howpublished}{\url{https://coinmarketcap.com/all/views/all/}}.
\newblock
\newblock
\shownote{[Online; accessed 19-August-2019].}


\bibitem[\protect\citeauthoryear{??}{dif}{2019}]%
        {differentaltcoins}
 \bibinfo{year}{2019}\natexlab{}.
\newblock \bibinfo{title}{{Altcoins vs. Tokens: What's the difference?}}
\newblock
  \bibinfo{howpublished}{\url{https://masterthecrypto.com/differences-between-cryptocurrency-coins-and-tokens/}}.
\newblock
\newblock
\shownote{[Online; accessed 30-August-2019].}


\bibitem[\protect\citeauthoryear{??}{Bit}{2019b}]%
        {Bitcoin}
 \bibinfo{year}{2019}\natexlab{b}.
\newblock \bibinfo{title}{{Bitcoin price, charts, market cap and other metrics
  | CoinMarketCap}}.
\newblock
  \bibinfo{howpublished}{\url{https://coinmarketcap.com/currencies/bitcoin/}}.
\newblock
\newblock
\shownote{[Online; accessed 19-August-2019].}


\bibitem[\protect\citeauthoryear{??}{Git}{2019}]%
        {Github_API}
 \bibinfo{year}{2019}\natexlab{}.
\newblock \bibinfo{title}{{github-api·GitHub Topics}}.
\newblock \bibinfo{howpublished}{\url{https://github.com/topics/github-api}}.
\newblock
\newblock
\shownote{[Online; accessed 9-September-2019].}


\bibitem[\protect\citeauthoryear{??}{per}{2019}]%
        {percentage}
 \bibinfo{year}{2019}\natexlab{}.
\newblock \bibinfo{title}{{Global Charts | CoinMarketCap}}.
\newblock \bibinfo{howpublished}{\url{https://coinmarketcap.com/charts/}}.
\newblock
\newblock
\shownote{[Online; accessed 19-August-2019].}


\bibitem[\protect\citeauthoryear{Ain, Butt, Anwar, Azam, and Maqbool}{Ain
  et~al\mbox{.}}{2019}]%
        {ain2019a}
\bibfield{author}{\bibinfo{person}{Qurat~Ul Ain}, \bibinfo{person}{Wasi~Haider
  Butt}, \bibinfo{person}{Muhammad~Waseem Anwar}, \bibinfo{person}{Farooque
  Azam}, {and} \bibinfo{person}{Bilal Maqbool}.}
  \bibinfo{year}{2019}\natexlab{}.
\newblock \showarticletitle{A Systematic Review on Code Clone Detection}.
\newblock \bibinfo{journal}{\emph{IEEE Access}}  \bibinfo{volume}{7}
  (\bibinfo{year}{2019}), \bibinfo{pages}{86121--86144}.
\newblock


\bibitem[\protect\citeauthoryear{Akram, Shi, Mumtaz, and Luo}{Akram
  et~al\mbox{.}}{2018}]%
        {DBLP:conf/compsac/AkramSML18}
\bibfield{author}{\bibinfo{person}{Junaid Akram}, \bibinfo{person}{Zhendong
  Shi}, \bibinfo{person}{Majid Mumtaz}, {and} \bibinfo{person}{Ping Luo}.}
  \bibinfo{year}{2018}\natexlab{}.
\newblock \showarticletitle{DroidCC: {A} Scalable Clone Detection Approach for
  Android Applications to Detect Similarity at Source Code Level}. In
  \bibinfo{booktitle}{\emph{2018 {IEEE} 42nd Annual Computer Software and
  Applications Conference, {COMPSAC} 2018, Tokyo, Japan, 23-27 July 2018,
  Volume 1}}. \bibinfo{pages}{100--105}.
\newblock


\bibitem[\protect\citeauthoryear{Alhazmi and Malaiya}{Alhazmi and
  Malaiya}{2005}]%
        {alhazmi2005quantitative}
\bibfield{author}{\bibinfo{person}{Omar~H Alhazmi} {and}
  \bibinfo{person}{Yashwant~K Malaiya}.} \bibinfo{year}{2005}\natexlab{}.
\newblock \showarticletitle{Quantitative vulnerability assessment of systems
  software}.
\newblock  (\bibinfo{year}{2005}), \bibinfo{pages}{615--620}.
\newblock


\bibitem[\protect\citeauthoryear{Apostolaki, Zohar, and Vanbever}{Apostolaki
  et~al\mbox{.}}{2017}]%
        {apostolaki2017hijacking}
\bibfield{author}{\bibinfo{person}{Maria Apostolaki}, \bibinfo{person}{Aviv
  Zohar}, {and} \bibinfo{person}{Laurent Vanbever}.}
  \bibinfo{year}{2017}\natexlab{}.
\newblock \showarticletitle{Hijacking Bitcoin: Routing Attacks on
  Cryptocurrencies}.
\newblock  (\bibinfo{year}{2017}), \bibinfo{pages}{375--392}.
\newblock


\bibitem[\protect\citeauthoryear{Azouvi, Maller, and Meiklejohn}{Azouvi
  et~al\mbox{.}}{2018}]%
        {DBLP:conf/fc/AzouviMM18}
\bibfield{author}{\bibinfo{person}{Sarah Azouvi}, \bibinfo{person}{Mary
  Maller}, {and} \bibinfo{person}{Sarah Meiklejohn}.}
  \bibinfo{year}{2018}\natexlab{}.
\newblock \showarticletitle{Egalitarian Society or Benevolent Dictatorship: The
  State of Cryptocurrency Governance}. In \bibinfo{booktitle}{\emph{Financial
  Cryptography and Data Security - {FC} 2018 International Workshops, BITCOIN,
  VOTING, and WTSC, Nieuwpoort, Cura{\c{c}}ao, March 2, 2018, Revised Selected
  Papers}}. \bibinfo{pages}{127--143}.
\newblock


\bibitem[\protect\citeauthoryear{Bartoletti, Carta, Cimoli, and
  Saia}{Bartoletti et~al\mbox{.}}{2017}]%
        {bartoletti2017dissecting}
\bibfield{author}{\bibinfo{person}{Massimo Bartoletti},
  \bibinfo{person}{Salvatore Carta}, \bibinfo{person}{Tiziana Cimoli}, {and}
  \bibinfo{person}{Roberto Saia}.} \bibinfo{year}{2017}\natexlab{}.
\newblock \showarticletitle{Dissecting Ponzi schemes on Ethereum:
  identification, analysis, and impact}.
\newblock \bibinfo{journal}{\emph{arXiv: Cryptography and Security}}
  (\bibinfo{year}{2017}).
\newblock


\bibitem[\protect\citeauthoryear{Biryukov, Khovratovich, and
  Pustogarov}{Biryukov et~al\mbox{.}}{2014}]%
        {biryukov2014deanonymisation}
\bibfield{author}{\bibinfo{person}{Alex Biryukov}, \bibinfo{person}{Dmitry
  Khovratovich}, {and} \bibinfo{person}{Ivan Pustogarov}.}
  \bibinfo{year}{2014}\natexlab{}.
\newblock \showarticletitle{Deanonymisation of Clients in Bitcoin P2P Network}.
\newblock  (\bibinfo{year}{2014}), \bibinfo{pages}{15--29}.
\newblock


\bibitem[\protect\citeauthoryear{Chen, Zheng, Cui, Ngai, Zheng, and Zhou}{Chen
  et~al\mbox{.}}{2018}]%
        {DBLP:conf/www/ChenZCNZZ18}
\bibfield{author}{\bibinfo{person}{Weili Chen}, \bibinfo{person}{Zibin Zheng},
  \bibinfo{person}{Jiahui Cui}, \bibinfo{person}{Edith C.~H. Ngai},
  \bibinfo{person}{Peilin Zheng}, {and} \bibinfo{person}{Yuren Zhou}.}
  \bibinfo{year}{2018}\natexlab{}.
\newblock \showarticletitle{Detecting Ponzi Schemes on Ethereum: Towards
  Healthier Blockchain Technology}. In \bibinfo{booktitle}{\emph{Proceedings of
  the 2018 World Wide Web Conference on World Wide Web, {WWW} 2018, Lyon,
  France, April 23-27, 2018}}. \bibinfo{pages}{1409--1418}.
\newblock


\bibitem[\protect\citeauthoryear{Chodarev, Pietrikova, and Kollar}{Chodarev
  et~al\mbox{.}}{2015}]%
        {chodarev2015haskell}
\bibfield{author}{\bibinfo{person}{Sergej Chodarev}, \bibinfo{person}{Emilia
  Pietrikova}, {and} \bibinfo{person}{Jan Kollar}.}
  \bibinfo{year}{2015}\natexlab{}.
\newblock \showarticletitle{Haskell clone detection using pattern comparing
  algorithm}.
\newblock  (\bibinfo{year}{2015}), \bibinfo{pages}{1--4}.
\newblock


\bibitem[\protect\citeauthoryear{Christin and Safavi{-}Naini}{Christin and
  Safavi{-}Naini}{2014}]%
        {DBLP:conf/fc/2014}
\bibfield{editor}{\bibinfo{person}{Nicolas Christin} {and}
  \bibinfo{person}{Reihaneh Safavi{-}Naini}} (Eds.).
  \bibinfo{year}{2014}\natexlab{}.
\newblock \bibinfo{booktitle}{\emph{Financial Cryptography and Data Security -
  18th International Conference, {FC} 2014, Christ Church, Barbados, March 3-7,
  2014, Revised Selected Papers}}. \bibinfo{series}{Lecture Notes in Computer
  Science}, Vol.~\bibinfo{volume}{8437}. \bibinfo{publisher}{Springer}.
\newblock


\bibitem[\protect\citeauthoryear{Crussell, Gibler, and Chen}{Crussell
  et~al\mbox{.}}{2015}]%
        {DBLP:journals/tmc/CrussellGC15}
\bibfield{author}{\bibinfo{person}{Jonathan Crussell}, \bibinfo{person}{Clint
  Gibler}, {and} \bibinfo{person}{Hao Chen}.} \bibinfo{year}{2015}\natexlab{}.
\newblock \showarticletitle{AnDarwin: Scalable Detection of Android Application
  Clones Based on Semantics}.
\newblock \bibinfo{journal}{\emph{{IEEE} Trans. Mob. Comput.}}
  \bibinfo{volume}{14}, \bibinfo{number}{10} (\bibinfo{year}{2015}),
  \bibinfo{pages}{2007--2019}.
\newblock


\bibitem[\protect\citeauthoryear{Fan, Liu, Luo, Chen, Tian, Zheng, and Liu}{Fan
  et~al\mbox{.}}{2018}]%
        {fan2018android}
\bibfield{author}{\bibinfo{person}{Ming Fan}, \bibinfo{person}{Jun Liu},
  \bibinfo{person}{Xiapu Luo}, \bibinfo{person}{Kai Chen},
  \bibinfo{person}{Zhenzhou Tian}, \bibinfo{person}{Qinghua Zheng}, {and}
  \bibinfo{person}{Ting Liu}.} \bibinfo{year}{2018}\natexlab{}.
\newblock \showarticletitle{Android malware familial classification and
  representative sample selection via frequent subgraph analysis}.
\newblock \bibinfo{journal}{\emph{IEEE TIFS}} \bibinfo{volume}{13},
  \bibinfo{number}{8} (\bibinfo{year}{2018}), \bibinfo{pages}{1890--1905}.
\newblock


\bibitem[\protect\citeauthoryear{Fan, Luo, Liu, Nong, Zheng, and Liu}{Fan
  et~al\mbox{.}}{2019a}]%
        {fan2019ctdroid}
\bibfield{author}{\bibinfo{person}{Ming Fan}, \bibinfo{person}{Xiapu Luo},
  \bibinfo{person}{Jun Liu}, \bibinfo{person}{Chunyin Nong},
  \bibinfo{person}{Qinghua Zheng}, {and} \bibinfo{person}{Ting Liu}.}
  \bibinfo{year}{2019}\natexlab{a}.
\newblock \showarticletitle{CTDroid: leveraging a corpus of technical blogs for
  android malware analysis}.
\newblock \bibinfo{journal}{\emph{IEEE Transactions on Reliability}}
  (\bibinfo{year}{2019}).
\newblock


\bibitem[\protect\citeauthoryear{Fan, Luo, Liu, Wang, Nong, Zheng, and Liu}{Fan
  et~al\mbox{.}}{2019b}]%
        {fan2019graph}
\bibfield{author}{\bibinfo{person}{Ming Fan}, \bibinfo{person}{Xiapu Luo},
  \bibinfo{person}{Jun Liu}, \bibinfo{person}{Meng Wang},
  \bibinfo{person}{Chunyin Nong}, \bibinfo{person}{Qinghua Zheng}, {and}
  \bibinfo{person}{Ting Liu}.} \bibinfo{year}{2019}\natexlab{b}.
\newblock \showarticletitle{Graph embedding based familial analysis of Android
  malware using unsupervised learning}. In
  \bibinfo{booktitle}{\emph{Proc.ICSE}}. IEEE, \bibinfo{pages}{771--782}.
\newblock


\bibitem[\protect\citeauthoryear{Hu, Zhang, Li, and Gu}{Hu
  et~al\mbox{.}}{2017}]%
        {DBLP:conf/iwpc/HuZLG17}
\bibfield{author}{\bibinfo{person}{Yikun Hu}, \bibinfo{person}{Yuanyuan Zhang},
  \bibinfo{person}{Juanru Li}, {and} \bibinfo{person}{Dawu Gu}.}
  \bibinfo{year}{2017}\natexlab{}.
\newblock \showarticletitle{Binary code clone detection across architectures
  and compiling configurations}. In \bibinfo{booktitle}{\emph{Proceedings of
  the 25th International Conference on Program Comprehension, {ICPC} 2017,
  Buenos Aires, Argentina, May 22-23, 2017}}. \bibinfo{pages}{88--98}.
\newblock


\bibitem[\protect\citeauthoryear{Jadon}{Jadon}{2016}]%
        {jadon2016code}
\bibfield{author}{\bibinfo{person}{Shruti Jadon}.}
  \bibinfo{year}{2016}\natexlab{}.
\newblock \showarticletitle{Code clones detection using machine learning
  technique: Support vector machine}.
\newblock  (\bibinfo{year}{2016}).
\newblock


\bibitem[\protect\citeauthoryear{Kamalpriya and Singh}{Kamalpriya and
  Singh}{2017}]%
        {DBLP:conf/iwsc/KamalpriyaS17}
\bibfield{author}{\bibinfo{person}{C.~M. Kamalpriya} {and}
  \bibinfo{person}{Paramvir Singh}.} \bibinfo{year}{2017}\natexlab{}.
\newblock \showarticletitle{Enhancing program dependency graph based clone
  detection using approximate subgraph matching}. In
  \bibinfo{booktitle}{\emph{11th {IEEE} International Workshop on Software
  Clones, {IWSC} 2017, Klagenfurt, Austria, February 21, 2017}}.
  \bibinfo{pages}{61--67}.
\newblock


\bibitem[\protect\citeauthoryear{Kappos, Yousaf, Maller, and Meiklejohn}{Kappos
  et~al\mbox{.}}{2018}]%
        {DBLP:conf/uss/KapposYMM18}
\bibfield{author}{\bibinfo{person}{George Kappos}, \bibinfo{person}{Haaroon
  Yousaf}, \bibinfo{person}{Mary Maller}, {and} \bibinfo{person}{Sarah
  Meiklejohn}.} \bibinfo{year}{2018}\natexlab{}.
\newblock \showarticletitle{An Empirical Analysis of Anonymity in Zcash}. In
  \bibinfo{booktitle}{\emph{27th {USENIX} Security Symposium, {USENIX} Security
  2018, Baltimore, MD, USA, August 15-17, 2018.}} \bibinfo{pages}{463--477}.
\newblock


\bibitem[\protect\citeauthoryear{Karp and Rabin}{Karp and Rabin}{1987}]%
        {karp1987efficient}
\bibfield{author}{\bibinfo{person}{Richard~M Karp} {and}
  \bibinfo{person}{Michael~O Rabin}.} \bibinfo{year}{1987}\natexlab{}.
\newblock \showarticletitle{Efficient randomized pattern-matching algorithms}.
\newblock \bibinfo{journal}{\emph{Ibm Journal of Research and Development}}
  \bibinfo{volume}{31}, \bibinfo{number}{2} (\bibinfo{year}{1987}),
  \bibinfo{pages}{249--260}.
\newblock


\bibitem[\protect\citeauthoryear{Kim, Woo, Lee, and Oh}{Kim
  et~al\mbox{.}}{2017}]%
        {kim2017vuddy}
\bibfield{author}{\bibinfo{person}{Seulbae Kim}, \bibinfo{person}{Seunghoon
  Woo}, \bibinfo{person}{Heejo Lee}, {and} \bibinfo{person}{Hakjoo Oh}.}
  \bibinfo{year}{2017}\natexlab{}.
\newblock \showarticletitle{VUDDY: A Scalable Approach for Vulnerable Code
  Clone Discovery}.
\newblock  (\bibinfo{year}{2017}), \bibinfo{pages}{595--614}.
\newblock


\bibitem[\protect\citeauthoryear{Liu, Wei, and Cao}{Liu et~al\mbox{.}}{2017}]%
        {liu2017vfdetect}
\bibfield{author}{\bibinfo{person}{Zhen Liu}, \bibinfo{person}{Qiang Wei},
  {and} \bibinfo{person}{Yan Cao}.} \bibinfo{year}{2017}\natexlab{}.
\newblock \showarticletitle{Vfdetect: A vulnerable code clone detection system
  based on vulnerability fingerprint}. In \bibinfo{booktitle}{\emph{2017 IEEE
  3rd Information Technology and Mechatronics Engineering Conference (ITOEC)}}.
  IEEE, \bibinfo{pages}{548--553}.
\newblock


\bibitem[\protect\citeauthoryear{Meiklejohn, Pomarole, Jordan, Levchenko,
  McCoy, Voelker, and Savage}{Meiklejohn et~al\mbox{.}}{2016}]%
        {DBLP:journals/cacm/MeiklejohnPJLMV16}
\bibfield{author}{\bibinfo{person}{Sarah Meiklejohn}, \bibinfo{person}{Marjori
  Pomarole}, \bibinfo{person}{Grant Jordan}, \bibinfo{person}{Kirill
  Levchenko}, \bibinfo{person}{Damon McCoy}, \bibinfo{person}{Geoffrey~M.
  Voelker}, {and} \bibinfo{person}{Stefan Savage}.}
  \bibinfo{year}{2016}\natexlab{}.
\newblock \showarticletitle{A fistful of Bitcoins: characterizing payments
  among men with no names}.
\newblock \bibinfo{journal}{\emph{Commun. {ACM}}} \bibinfo{volume}{59},
  \bibinfo{number}{4} (\bibinfo{year}{2016}), \bibinfo{pages}{86--93}.
\newblock


\bibitem[\protect\citeauthoryear{Miller, M{\"{o}}ser, Lee, and
  Narayanan}{Miller et~al\mbox{.}}{2017}]%
        {DBLP:journals/corr/MillerMLN17}
\bibfield{author}{\bibinfo{person}{Andrew Miller}, \bibinfo{person}{Malte
  M{\"{o}}ser}, \bibinfo{person}{Kevin Lee}, {and} \bibinfo{person}{Arvind
  Narayanan}.} \bibinfo{year}{2017}\natexlab{}.
\newblock \showarticletitle{An Empirical Analysis of Linkability in the Monero
  Blockchain}.
\newblock \bibinfo{journal}{\emph{CoRR}}  \bibinfo{volume}{abs/1704.04299}
  (\bibinfo{year}{2017}).
\newblock


\bibitem[\protect\citeauthoryear{Misu and Sakib}{Misu and Sakib}{2017}]%
        {DBLP:conf/apsec/MisuS17}
\bibfield{author}{\bibinfo{person}{Md~Rakib~Hossain Misu} {and}
  \bibinfo{person}{Kazi Sakib}.} \bibinfo{year}{2017}\natexlab{}.
\newblock \showarticletitle{Interface Driven Code Clone Detection}. In
  \bibinfo{booktitle}{\emph{24th Asia-Pacific Software Engineering Conference,
  {APSEC} 2017, Nanjing, China, December 4-8, 2017}}.
  \bibinfo{pages}{747--748}.
\newblock


\bibitem[\protect\citeauthoryear{Misu, Satter, and Sakib}{Misu
  et~al\mbox{.}}{2017}]%
        {DBLP:conf/apsec/MisuSS17}
\bibfield{author}{\bibinfo{person}{Md~Rakib~Hossain Misu},
  \bibinfo{person}{Abdus Satter}, {and} \bibinfo{person}{Kazi Sakib}.}
  \bibinfo{year}{2017}\natexlab{}.
\newblock \showarticletitle{An Exploratory Study on Interface Similarities in
  Code Clones}. In \bibinfo{booktitle}{\emph{24th Asia-Pacific Software
  Engineering Conference Workshops, {APSEC} Workshops 2017, Nanjing, China,
  December 4-8, 2017}}. \bibinfo{pages}{126--133}.
\newblock


\bibitem[\protect\citeauthoryear{Nakamura, Choi, Yoshida, Haruna, and
  Inoue}{Nakamura et~al\mbox{.}}{2016}]%
        {DBLP:conf/wcre/NakamuraCYHI16}
\bibfield{author}{\bibinfo{person}{Yuta Nakamura}, \bibinfo{person}{Eunjong
  Choi}, \bibinfo{person}{Norihiro Yoshida}, \bibinfo{person}{Shusuke Haruna},
  {and} \bibinfo{person}{Katsuro Inoue}.} \bibinfo{year}{2016}\natexlab{}.
\newblock \showarticletitle{Towards Detection and Analysis of Interlanguage
  Clones for Multilingual Web Applications}. In \bibinfo{booktitle}{\emph{10th
  International Workshop on Software Clones, IWSC@SANER 2016, Osaka, Japan,
  March 15, 2016}}. \bibinfo{pages}{17--18}.
\newblock


\bibitem[\protect\citeauthoryear{Newman, Sage, Collard, Alomari, and
  Maletic}{Newman et~al\mbox{.}}{2016}]%
        {DBLP:conf/icse/NewmanSCAM16}
\bibfield{author}{\bibinfo{person}{Christian~D. Newman},
  \bibinfo{person}{Tessandra Sage}, \bibinfo{person}{Michael~L. Collard},
  \bibinfo{person}{Hakam~W. Alomari}, {and} \bibinfo{person}{Jonathan~I.
  Maletic}.} \bibinfo{year}{2016}\natexlab{}.
\newblock \showarticletitle{srcSlice: a tool for efficient static forward
  slicing}. In \bibinfo{booktitle}{\emph{Proceedings of the 38th International
  Conference on Software Engineering, {ICSE} 2016, Austin, TX, USA, May 14-22,
  2016 - Companion Volume}}. \bibinfo{pages}{621--624}.
\newblock


\bibitem[\protect\citeauthoryear{Nishi and Damevski}{Nishi and
  Damevski}{2018}]%
        {nishi2018scalable}
\bibfield{author}{\bibinfo{person}{Manziba~Akanda Nishi} {and}
  \bibinfo{person}{Kostadin Damevski}.} \bibinfo{year}{2018}\natexlab{}.
\newblock \showarticletitle{Scalable code clone detection and search based on
  adaptive prefix filtering}.
\newblock \bibinfo{journal}{\emph{Journal of Systems and Software}}
  \bibinfo{volume}{137} (\bibinfo{year}{2018}), \bibinfo{pages}{130--142}.
\newblock


\bibitem[\protect\citeauthoryear{Pati, Kumar, Manjhi, and Shukla}{Pati
  et~al\mbox{.}}{2017}]%
        {pati2017a}
\bibfield{author}{\bibinfo{person}{Jayadeep Pati}, \bibinfo{person}{Babloo
  Kumar}, \bibinfo{person}{Devesh Manjhi}, {and} \bibinfo{person}{K~K Shukla}.}
  \bibinfo{year}{2017}\natexlab{}.
\newblock \showarticletitle{A Comparison Among ARIMA, BP-NN, and MOGA-NN for
  Software Clone Evolution Prediction}.
\newblock \bibinfo{journal}{\emph{IEEE Access}}  \bibinfo{volume}{5}
  (\bibinfo{year}{2017}), \bibinfo{pages}{11841--11851}.
\newblock


\bibitem[\protect\citeauthoryear{Ragkhitwetsagul and Krinke}{Ragkhitwetsagul
  and Krinke}{2017}]%
        {ragkhitwetsagul2017using}
\bibfield{author}{\bibinfo{person}{Chaiyong Ragkhitwetsagul} {and}
  \bibinfo{person}{Jens Krinke}.} \bibinfo{year}{2017}\natexlab{}.
\newblock \showarticletitle{Using compilation/decompilation to enhance clone
  detection}.
\newblock  (\bibinfo{year}{2017}), \bibinfo{pages}{1--7}.
\newblock


\bibitem[\protect\citeauthoryear{Ragkhitwetsagul, Krinke, and
  Marnette}{Ragkhitwetsagul et~al\mbox{.}}{2018}]%
        {DBLP:conf/iwsc/Ragkhitwetsagul18}
\bibfield{author}{\bibinfo{person}{Chaiyong Ragkhitwetsagul},
  \bibinfo{person}{Jens Krinke}, {and} \bibinfo{person}{Bruno Marnette}.}
  \bibinfo{year}{2018}\natexlab{}.
\newblock \showarticletitle{A picture is worth a thousand words: Code clone
  detection based on image similarity}. In \bibinfo{booktitle}{\emph{12th
  {IEEE} International Workshop on Software Clones, {IWSC} 2018, Campobasso,
  Italy, March 20, 2018}}. \bibinfo{pages}{44--50}.
\newblock


\bibitem[\protect\citeauthoryear{Reibel, Yousaf, and Meiklejohn}{Reibel
  et~al\mbox{.}}{2019}]%
        {DBLP:conf/fc/ReibelYM19}
\bibfield{author}{\bibinfo{person}{Pierre Reibel}, \bibinfo{person}{Haaroon
  Yousaf}, {and} \bibinfo{person}{Sarah Meiklejohn}.}
  \bibinfo{year}{2019}\natexlab{}.
\newblock \showarticletitle{Short Paper: An Exploration of Code Diversity in
  the Cryptocurrency Landscape}. In \bibinfo{booktitle}{\emph{Financial
  Cryptography and Data Security - 23rd International Conference, {FC} 2019,
  Frigate Bay, St. Kitts and Nevis, February 18-22, 2019, Revised Selected
  Papers}}. \bibinfo{pages}{73--83}.
\newblock


\bibitem[\protect\citeauthoryear{Reid and Harrigan}{Reid and Harrigan}{2011}]%
        {DBLP:conf/socialcom/ReidH11}
\bibfield{author}{\bibinfo{person}{Fergal Reid} {and} \bibinfo{person}{Martin
  Harrigan}.} \bibinfo{year}{2011}\natexlab{}.
\newblock \showarticletitle{An Analysis of Anonymity in the Bitcoin System}. In
  \bibinfo{booktitle}{\emph{PASSAT/SocialCom 2011, Privacy, Security, Risk and
  Trust (PASSAT), 2011 {IEEE} Third International Conference on and 2011 {IEEE}
  Third International Conference on Social Computing (SocialCom), Boston, MA,
  USA, 9-11 Oct., 2011}}. \bibinfo{pages}{1318--1326}.
\newblock


\bibitem[\protect\citeauthoryear{Roopam and Singh}{Roopam and Singh}{2017}]%
        {roopam2017to}
\bibfield{author}{\bibinfo{person}{Roopam} {and} \bibinfo{person}{Gurpreet
  Singh}.} \bibinfo{year}{2017}\natexlab{}.
\newblock \showarticletitle{To enhance the code clone detection algorithm by
  using hybrid approach for detection of code clones}.
\newblock  (\bibinfo{year}{2017}).
\newblock


\bibitem[\protect\citeauthoryear{Ruffing, Thyagarajan, Ronge, and
  Schroder}{Ruffing et~al\mbox{.}}{2018}]%
        {ruffing2018burning}
\bibfield{author}{\bibinfo{person}{Tim Ruffing}, \bibinfo{person}{Sri
  Aravinda~Krishnan Thyagarajan}, \bibinfo{person}{Viktoria Ronge}, {and}
  \bibinfo{person}{Dominique Schroder}.} \bibinfo{year}{2018}\natexlab{}.
\newblock \showarticletitle{Burning Zerocoins for Fun and for Profit: A
  Cryptographic Denial-of-Spending Attack on the Zerocoin Protocol.}
\newblock \bibinfo{journal}{\emph{IACR Cryptology ePrint Archive}}
  \bibinfo{volume}{2018} (\bibinfo{year}{2018}), \bibinfo{pages}{612}.
\newblock


\bibitem[\protect\citeauthoryear{Sabi, Higo, and Kusumoto}{Sabi
  et~al\mbox{.}}{2017}]%
        {DBLP:conf/iwsc/SabiHK17}
\bibfield{author}{\bibinfo{person}{Yusuke Sabi}, \bibinfo{person}{Yoshiki
  Higo}, {and} \bibinfo{person}{Shinji Kusumoto}.}
  \bibinfo{year}{2017}\natexlab{}.
\newblock \showarticletitle{Rearranging the order of program statements for
  code clone detection}. In \bibinfo{booktitle}{\emph{11th {IEEE} International
  Workshop on Software Clones, {IWSC} 2017, Klagenfurt, Austria, February 21,
  2017}}. \bibinfo{pages}{15--21}.
\newblock


\bibitem[\protect\citeauthoryear{Sajnani, Saini, Svajlenko, Roy, and
  Lopes}{Sajnani et~al\mbox{.}}{2016}]%
        {sajnani2016sourcerercc}
\bibfield{author}{\bibinfo{person}{Hitesh Sajnani}, \bibinfo{person}{Vaibhav
  Saini}, \bibinfo{person}{Jeffrey Svajlenko}, \bibinfo{person}{Chanchal~K
  Roy}, {and} \bibinfo{person}{Cristina~Videira Lopes}.}
  \bibinfo{year}{2016}\natexlab{}.
\newblock \showarticletitle{SourcererCC: scaling code clone detection to
  big-code}.
\newblock \bibinfo{journal}{\emph{international conference on software
  engineering}} (\bibinfo{year}{2016}), \bibinfo{pages}{1157--1168}.
\newblock


\bibitem[\protect\citeauthoryear{Semura, Yoshida, Choi, and Inoue}{Semura
  et~al\mbox{.}}{2017}]%
        {DBLP:conf/apsec/SemuraYCI17}
\bibfield{author}{\bibinfo{person}{Yuichi Semura}, \bibinfo{person}{Norihiro
  Yoshida}, \bibinfo{person}{Eunjong Choi}, {and} \bibinfo{person}{Katsuro
  Inoue}.} \bibinfo{year}{2017}\natexlab{}.
\newblock \showarticletitle{CCFinderSW: Clone Detection Tool with Flexible
  Multilingual Tokenization}. In \bibinfo{booktitle}{\emph{24th Asia-Pacific
  Software Engineering Conference, {APSEC} 2017, Nanjing, China, December 4-8,
  2017}}. \bibinfo{pages}{654--659}.
\newblock


\bibitem[\protect\citeauthoryear{Sheneamer, Roy, and Kalita}{Sheneamer
  et~al\mbox{.}}{2018}]%
        {DBLP:journals/eswa/SheneamerRK18}
\bibfield{author}{\bibinfo{person}{Abdullah Sheneamer}, \bibinfo{person}{Swarup
  Roy}, {and} \bibinfo{person}{Jugal Kalita}.} \bibinfo{year}{2018}\natexlab{}.
\newblock \showarticletitle{A detection framework for semantic code clones and
  obfuscated code}.
\newblock \bibinfo{journal}{\emph{Expert Syst. Appl.}}  \bibinfo{volume}{97}
  (\bibinfo{year}{2018}), \bibinfo{pages}{405--420}.
\newblock


\bibitem[\protect\citeauthoryear{Spagnuolo, Maggi, and Zanero}{Spagnuolo
  et~al\mbox{.}}{2014}]%
        {DBLP:conf/fc/SpagnuoloMZ14}
\bibfield{author}{\bibinfo{person}{Michele Spagnuolo},
  \bibinfo{person}{Federico Maggi}, {and} \bibinfo{person}{Stefano Zanero}.}
  \bibinfo{year}{2014}\natexlab{}.
\newblock \showarticletitle{BitIodine: Extracting Intelligence from the Bitcoin
  Network}. In \bibinfo{booktitle}{\emph{Financial Cryptography and Data
  Security - 18th International Conference, {FC} 2014, Christ Church, Barbados,
  March 3-7, 2014, Revised Selected Papers}}. \bibinfo{pages}{457--468}.
\newblock


\bibitem[\protect\citeauthoryear{Svajlenko and Roy}{Svajlenko and Roy}{2017a}]%
        {DBLP:conf/icse/SvajlenkoR17a}
\bibfield{author}{\bibinfo{person}{Jeffrey Svajlenko} {and}
  \bibinfo{person}{Chanchal~K. Roy}.} \bibinfo{year}{2017}\natexlab{a}.
\newblock \showarticletitle{CloneWorks: a fast and flexible large-scale
  near-miss clone detection tool}. In \bibinfo{booktitle}{\emph{Proceedings of
  the 39th International Conference on Software Engineering, {ICSE} 2017,
  Buenos Aires, Argentina, May 20-28, 2017 - Companion Volume}}.
  \bibinfo{pages}{177--179}.
\newblock


\bibitem[\protect\citeauthoryear{Svajlenko and Roy}{Svajlenko and Roy}{2017b}]%
        {DBLP:conf/icse/SvajlenkoR17}
\bibfield{author}{\bibinfo{person}{Jeffrey Svajlenko} {and}
  \bibinfo{person}{Chanchal~Kumar Roy}.} \bibinfo{year}{2017}\natexlab{b}.
\newblock \showarticletitle{Fast and flexible large-scale clone detection with
  CloneWorks}. In \bibinfo{booktitle}{\emph{Proceedings of the 39th
  International Conference on Software Engineering, {ICSE} 2017, Buenos Aires,
  Argentina, May 20-28, 2017 - Companion Volume}}. \bibinfo{pages}{27--30}.
\newblock


\bibitem[\protect\citeauthoryear{Tekchandani, Bhatia, and Singh}{Tekchandani
  et~al\mbox{.}}{2017}]%
        {DBLP:journals/cee/TekchandaniBS17}
\bibfield{author}{\bibinfo{person}{Rajkumar Tekchandani},
  \bibinfo{person}{Rajesh~Kumar Bhatia}, {and} \bibinfo{person}{Maninder
  Singh}.} \bibinfo{year}{2017}\natexlab{}.
\newblock \showarticletitle{Code clone genealogy detection on e-health system
  using Hadoop}.
\newblock \bibinfo{journal}{\emph{Computers {\&} Electrical Engineering}}
  \bibinfo{volume}{61} (\bibinfo{year}{2017}), \bibinfo{pages}{15--30}.
\newblock


\bibitem[\protect\citeauthoryear{Thaller, Ramler, Pichler, and Egyed}{Thaller
  et~al\mbox{.}}{2017}]%
        {DBLP:conf/etfa/ThallerRPE17}
\bibfield{author}{\bibinfo{person}{Hannes Thaller}, \bibinfo{person}{Rudolf
  Ramler}, \bibinfo{person}{Josef Pichler}, {and} \bibinfo{person}{Alexander
  Egyed}.} \bibinfo{year}{2017}\natexlab{}.
\newblock \showarticletitle{Exploring code clones in programmable logic
  controller software}. In \bibinfo{booktitle}{\emph{22nd {IEEE} International
  Conference on Emerging Technologies and Factory Automation, {ETFA} 2017,
  Limassol, Cyprus, September 12-15, 2017}}. \bibinfo{pages}{1--8}.
\newblock


\bibitem[\protect\citeauthoryear{Vasek, Thornton, and Moore}{Vasek
  et~al\mbox{.}}{2014}]%
        {DBLP:conf/fc/VasekTM14}
\bibfield{author}{\bibinfo{person}{Marie Vasek}, \bibinfo{person}{Micah
  Thornton}, {and} \bibinfo{person}{Tyler Moore}.}
  \bibinfo{year}{2014}\natexlab{}.
\newblock \showarticletitle{Empirical Analysis of Denial-of-Service Attacks in
  the Bitcoin Ecosystem}. In \bibinfo{booktitle}{\emph{Financial Cryptography
  and Data Security - {FC} 2014 Workshops, {BITCOIN} and {WAHC} 2014, Christ
  Church, Barbados, March 7, 2014, Revised Selected Papers}}.
  \bibinfo{pages}{57--71}.
\newblock


\bibitem[\protect\citeauthoryear{Vislavski, Rakic, Cardozo, and
  Budimac}{Vislavski et~al\mbox{.}}{2018}]%
        {DBLP:conf/wcre/VislavskiRCB18}
\bibfield{author}{\bibinfo{person}{Tijana Vislavski}, \bibinfo{person}{Gordana
  Rakic}, \bibinfo{person}{Nicol{\'{a}}s Cardozo}, {and} \bibinfo{person}{Zoran
  Budimac}.} \bibinfo{year}{2018}\natexlab{}.
\newblock \showarticletitle{{LICCA:} {A} tool for cross-language clone
  detection}. In \bibinfo{booktitle}{\emph{25th International Conference on
  Software Analysis, Evolution and Reengineering, {SANER} 2018, Campobasso,
  Italy, March 20-23, 2018}}. \bibinfo{pages}{512--516}.
\newblock


\bibitem[\protect\citeauthoryear{Wang, Wang, and Xu}{Wang
  et~al\mbox{.}}{2017}]%
        {DBLP:conf/apsec/WangWX17}
\bibfield{author}{\bibinfo{person}{Min Wang}, \bibinfo{person}{Pengcheng Wang},
  {and} \bibinfo{person}{Yun Xu}.} \bibinfo{year}{2017}\natexlab{}.
\newblock \showarticletitle{CCSharp: An Efficient Three-Phase Code Clone
  Detector Using Modified PDGs}. In \bibinfo{booktitle}{\emph{24th Asia-Pacific
  Software Engineering Conference, {APSEC} 2017, Nanjing, China, December 4-8,
  2017}}. \bibinfo{pages}{100--109}.
\newblock


\bibitem[\protect\citeauthoryear{Wang, Svajlenko, Wu, Xu, and Roy}{Wang
  et~al\mbox{.}}{2018}]%
        {DBLP:conf/icse/WangSWXR18}
\bibfield{author}{\bibinfo{person}{Pengcheng Wang}, \bibinfo{person}{Jeffrey
  Svajlenko}, \bibinfo{person}{Yanzhao Wu}, \bibinfo{person}{Yun Xu}, {and}
  \bibinfo{person}{Chanchal~K. Roy}.} \bibinfo{year}{2018}\natexlab{}.
\newblock \showarticletitle{CCAligner: a token based large-gap clone detector}.
  In \bibinfo{booktitle}{\emph{Proceedings of the 40th International Conference
  on Software Engineering, {ICSE} 2018, Gothenburg, Sweden, May 27 - June 03,
  2018}}. \bibinfo{pages}{1066--1077}.
\newblock


\bibitem[\protect\citeauthoryear{Xue, Venkataramani, and Lan}{Xue
  et~al\mbox{.}}{2018}]%
        {DBLP:conf/pldi/XueVL18}
\bibfield{author}{\bibinfo{person}{Hongfa Xue}, \bibinfo{person}{Guru
  Venkataramani}, {and} \bibinfo{person}{Tian Lan}.}
  \bibinfo{year}{2018}\natexlab{}.
\newblock \showarticletitle{Clone-hunter: accelerated bound checks elimination
  via binary code clone detection}. In \bibinfo{booktitle}{\emph{Proceedings of
  the 2nd {ACM} {SIGPLAN} International Workshop on Machine Learning and
  Programming Languages, MAPL@PLDI 2018, Philadelphia, PA, USA, June 18-22,
  2018}}. \bibinfo{pages}{11--19}.
\newblock


\bibitem[\protect\citeauthoryear{Yang, Ren, Chen, and Jiang}{Yang
  et~al\mbox{.}}{2018}]%
        {yang2018structural}
\bibfield{author}{\bibinfo{person}{Yanming Yang}, \bibinfo{person}{Zhilei Ren},
  \bibinfo{person}{Xin Chen}, {and} \bibinfo{person}{He Jiang}.}
  \bibinfo{year}{2018}\natexlab{}.
\newblock \showarticletitle{Structural Function Based Code Clone Detection
  Using a New Hybrid Technique}.
\newblock  (\bibinfo{year}{2018}), \bibinfo{pages}{286--291}.
\newblock


\bibitem[\protect\citeauthoryear{Yu, Wang, Wu, Yang, Wang, Yang, and Yan}{Yu
  et~al\mbox{.}}{2017}]%
        {yu2017detecting}
\bibfield{author}{\bibinfo{person}{Dongjin Yu}, \bibinfo{person}{Jie Wang},
  \bibinfo{person}{Qing Wu}, \bibinfo{person}{Jiazha Yang},
  \bibinfo{person}{Jiaojiao Wang}, \bibinfo{person}{Wei Yang}, {and}
  \bibinfo{person}{Wei Yan}.} \bibinfo{year}{2017}\natexlab{}.
\newblock \showarticletitle{Detecting Java Code Clones with Multi-granularities
  Based on Bytecode}.
\newblock  (\bibinfo{year}{2017}), \bibinfo{pages}{317--326}.
\newblock


\bibitem[\protect\citeauthoryear{Yuki, Higo, and Kusumoto}{Yuki
  et~al\mbox{.}}{2017}]%
        {yuki2017a}
\bibfield{author}{\bibinfo{person}{Yusuke Yuki}, \bibinfo{person}{Yoshiki
  Higo}, {and} \bibinfo{person}{Shinji Kusumoto}.}
  \bibinfo{year}{2017}\natexlab{}.
\newblock \showarticletitle{A technique to detect multi-grained code clones}.
\newblock  (\bibinfo{year}{2017}), \bibinfo{pages}{54--60}.
\newblock


\bibitem[\protect\citeauthoryear{Řican}{Řican}{2012}]%
        {historical}
\bibfield{author}{\bibinfo{person}{Oldřich Řican}.}
  \bibinfo{year}{2012}\natexlab{}.
\newblock \showarticletitle{Historical Biogeography of Neotropical Freshwater
  Fishes}.
\newblock \bibinfo{journal}{\emph{Systematic Biology}} \bibinfo{volume}{61},
  \bibinfo{number}{3} (\bibinfo{year}{2012}), \bibinfo{pages}{543--545}.
\newblock


\end{thebibliography}
\end{document}